\documentclass[twocolumn,showpacs,showkeys,preprintnumbers,amsmath,amssymb,aps,nofootXinbib,floatfix]{revtex4}
\usepackage{color}
\usepackage{indentfirst}
\usepackage{graphicx}
\usepackage{multirow}
      \def\di{\displaystyle}
      
      \def\bS{{\bf S}}
      \def\bl{{\bf l}}
      \def\bp{{\bf p}}
      
      \def\br{{\bf r}}

      \def\F{{\cal F}}
      \def\H{{\cal H}}

      \def\L{{\cal L}}
      \def\M{{\cal M}}
      
      \def\P{{\cal P}}
      \def\R{{\cal R}}

      \def\eq{{\rm eq}} 
      \def\Q{{Q_{20}}}   \def\bQ{{\bar{Q}_{20}}}    
      \def\Qx{{Q_{00}}}  \def\bQx{{\bar{Q}_{00}}}
      \def\e{{\rm e}}

      \def\u{{{\uparrow\downarrow}}}
      \def\d{{{\downarrow\uparrow}}}
      \def\uu{{{\uparrow\uparrow}}}
      \def\dd{{{\downarrow\downarrow}}}

\begin{document}

\title{Electric $1^+$ state below nuclear scissors}

\author{ E.B. Balbutsev\email{balbuts@theor.jinr.ru}, 
I.V. Molodtsova\email{molod@theor.jinr.ru}
}
\affiliation{Bogoliubov Laboratory of Theoretical Physics, Joint Institute for Nuclear Research, 141980 Dubna, Russia}

\begin{abstract}

The solution of time dependent Hartree-Fock-Bogoliubov 
equations by the Wigner function moments method
predicts four low-lying $1^+$ states. Three of them are known as various scissors modes.
Fourth state is disposed below all scissors modes and has the electrical nature.
It is found that it represents one of three branches of 
$2^+$ state which can exist in spherical nuclei and which is split 
in deformed nuclei. It is discovered, that the antiferromagnetic properties of
nuclei lead to the splitting of $2^+$ states already at the zero deformation.

\end{abstract}


\maketitle

\section{Introduction}\label{I}

The theoretical interpretation of collective nuclear dynamics, realized through electromagnetic transitions 
in the low-energy region, is one of the most interesting topics in nuclear structure physics. 
The method of Wigner function moments (WFM) or phase space moments turned out to be very convenient to describe the collective excitations of nuclei.
On the one hand, this is a purely microscopic approach, since it is based on the non-stationary Hartree-Fock-Bogolyubov equation. 
On the other hand, the method works with average values (moments) of operators,
which have a transparent physical meaning, and their dynamics is directly related to the processes under consideration. 
Thus, a natural bridge is thrown over with a macroscopic description.
All this makes the WFM method an ideal instrument for describing the main characteristics (energies and excitation probabilities) of collective excitations.

The description of the collective motion of nuclei in terms of WFM was first proposed in 1981~\cite{Ba81}. 
It has been successfully applied to study giant multipole resonances and low-lying collective modes of rotating and non-rotating nuclei 
with different realistic forces~\cite{Ba91,BaMo94}.
The catalyst for further progression along this path of research
was the understanding of the conceptual closeness of the WFM method and the variances-covariances approach 
developed by Peter Schuck~\cite{Sc84}. This formed the basis for joining efforts in this common direction of research. 
Initially, the method was applied to the study of large amplitude motion. 
Its Fourier analysis gave a lot of information about the variety of multiphonon states of nuclei~\cite{BaSc97,BaSc99}.

A significant point in the study of the nature of collective vibrations was the idea of orbital scissors,
first reported by R. Hilton in 1976 at the conference on nuclear structure in Dubna~\cite{Hilt}. 
This idea was subsequently developed in the works of Suzuki and Rowe~\cite{Suzuki}, Lo~Iudice and Palumbo~\cite{LoIP78} and other authors.
The first experimental detection of the nuclear scissors in $^{156}$Gd 
by the Darmstadt group~\cite{Bohle} has initiated a cascade of experimental 
and theoretical studies. An exhaustive review of the subject
is given in the paper~\cite{Heyd} containing about 400 references.

The detailed analysis of the scissors mode in the framework of a solvable model (harmonic oscillator with quadrupole-quadrupole residual interaction) 
was given in~\cite{BaSc}. These  investigations have shown
that in the small-amplitude approximation, already the minimal set of collective variables, i.e. phase space moments up to quadratic order, 
is sufficient to reproduce the most important property of the scissors mode: 
its inevitable coexistence with the isovector Giant Quadrupole Resonance (GQR) implying a deformation of the Fermi surface.
Additionally, this simple model, which allows one to obtain exact analytical solutions, 
proved to be a good basis for studying the interrelation between WFM method and Random Phase Approximation (RPA).  
It was shown that these two methods give identical formulae for eigenfrequencies and transition probabilities of all collective excitations of the model, 
including the scissors mode. On the whole, it can be concluded that 
the second order moment equations yield an optimally coarse grained image of the full QRPA spectrum~\cite{BaSc06,Ann}.

The simple model provided a lot of useful information about collective motion, 
correctly conveyed the observed tendences for scissors mode, and helped to better understand the potential capability  of the method of moments 
and its relationship to microscopics.
However, the energy of the scissors mode turned out to be too low, and the values of $B(M1)$ were too large compared to the experimental data.
Accounting for pair correlations significantly improved agreement with experiment, but did not completely solve the problem~\cite{Malov,Malov1}.
It became clear that the spin-orbit interaction must be included in the consideration.
In the paper~\cite{BaMo} the WFM method was generalized
 to solve the TDHF equations including spin
dynamics. The most remarkable result was the prediction of a new type
of nuclear collective motion: rotational oscillations of "spin up"
nucleons with respect to "spin down" nucleons (the spin scissors mode). 
A generalization of the WFM method which takes into account spin degrees
of freedom and pair correlations simultaneously was outlined in~\cite{BaMoPRC2}, where the Time Dependent Hartree-Fock-Bogoliubov (TDHFB)
equations were considered. As a result the agreement between theory and
experiment in the description of nuclear scissors modes was improved considerably.
Furthermore, after taking into account the isovector-isoscalar coupling~\cite{BaMoPRC22} one more magnetic mode (third type of scissors) emerged.
Actually, the possible existence of three scissors motions is 
easily explained by  combinatoric
consideration -- there are only three ways to divide the four different kinds 
of objects (spin up and spin down protons and neutrons in our case) into two pairs. 
The three types  of scissors modes can be approximately classified as isovector 
spin-scalar (conventional), isovector spin-vector and
isoscalar spin-vector.

All these results have been obtained through many years of exciting and fruitful collaboration with Peter Schuck, 
whose intellectual contribution to the development of the WFM method is invaluable.

The WFM based calculations also predict low-lying
excitation with large $B(E2)$ and negligible $B(M1)$ values, which is disposed just below all scissors modes. 
Until now, we have left this topic out of the discussion.
The focus of the present work is an analysis of the nature of this state.

\section{TDHFB equations and WFM equations of motion}\label{II}

\hspace{5mm} The TDHFB equations in matrix formulation \cite{Solov,Ring} are
\begin{equation}
i\hbar\dot\R=[\H,\R]
\label{tHFB}
\end{equation}
with
\begin{equation}
\R={\hat\rho\qquad-\hat\kappa\choose-\hat\kappa^{\dagger}\;\;1-\hat\rho^*},
\quad\H={\hat
h\quad\;\;\hat\Delta\choose\hat\Delta^{\dagger}\quad-\hat h^*}
\end{equation}
The normal density matrix $\hat \rho$ and Hamiltonian $\hat h$ are
hermitian whereas the anomalous density $\hat\kappa$ and the pairing
gap $\hat\Delta$ are skew symmetric: $\hat\kappa^{\dagger}=-\hat\kappa^*$, 
$\hat\Delta^{\dagger}=-\hat\Delta^*$.
The detailed form of the TDHFB equations is
\begin{eqnarray}
&& i\hbar\dot{\hat\rho} =\hat h\hat\rho -\hat\rho\hat h
-\hat\Delta \hat\kappa ^{\dagger}+\hat\kappa \hat\Delta^\dagger,
\nonumber\\
&&-i\hbar\dot{\hat\rho}^*=\hat h^*\hat\rho ^*-\hat\rho ^*\hat h^*
-\hat\Delta^\dagger\hat\kappa +\hat\kappa^\dagger\hat\Delta ,
\nonumber\\
&&-i\hbar\dot{\hat\kappa} =-\hat h\hat\kappa -\hat\kappa \hat h^*+\hat\Delta
-\hat\Delta \hat\rho ^*-\hat\rho \hat\Delta ,
\nonumber\\
&&-i\hbar\dot{\hat\kappa}^\dagger=\hat h^*\hat\kappa^\dagger
+\hat\kappa^\dagger\hat h-\hat\Delta^\dagger
+\hat\Delta^\dagger\hat\rho +\hat\rho^*\hat\Delta^\dagger .
\label{HFB}
\end{eqnarray}
Let us consider their matrix form in coordinate 
space keeping all spin indices $s, s'$:
$\langle \br,s|\hat\rho|\br',s'\rangle$, 
$\langle \br,s|\hat\kappa|\br',s'\rangle$, etc.
 We do not specify the isospin indices in order to make
formulae more transparent. 
After introduction of the more compact notation
$\langle \br,s|\hat X|\br',s'\rangle =X_{rr'}^{ss'}$
the set of equations (\ref{HFB}) with specified spin indices reads

\begin{widetext}

\begin{eqnarray}
&&i\hbar\dot{\rho}_{rr''}^{\uparrow\uparrow} =
\int\!d^3r'(
 h_{rr'}^{\uparrow\uparrow}\rho_{r'r''}^{\uparrow\uparrow} 
-\rho_{rr'}^{\uparrow\uparrow} h_{r'r''}^{\uparrow\uparrow}
+ h_{rr'}^{\uparrow\downarrow}\rho_{r'r''}^{\downarrow\uparrow} 
-\rho_{rr'}^{\uparrow\downarrow} h_{r'r''}^{\downarrow\uparrow}
-\Delta_{rr'}^{\uparrow\downarrow}{\kappa^{\dagger}}_{r'r''}^{\downarrow\uparrow}
+\kappa_{rr'}^{\uparrow\downarrow}{\Delta^{\dagger}}_{r'r''}^{\downarrow\uparrow}),
\nonumber\\
&&i\hbar\dot{\rho}_{rr''}^{\uparrow\downarrow} =
\int\!d^3r'(
 h_{rr'}^{\uparrow\uparrow}\rho_{r'r''}^{\uparrow\downarrow} 
-\rho_{rr'}^{\uparrow\uparrow} h_{r'r''}^{\uparrow\downarrow}
+ h_{rr'}^{\uparrow\downarrow}\rho_{r'r''}^{\downarrow\downarrow} 
-\rho_{rr'}^{\uparrow\downarrow} h_{r'r''}^{\downarrow\downarrow}),
\nonumber\\
&&i\hbar\dot{\rho}_{rr''}^{\downarrow\uparrow} =
\int\!d^3r'(
 h_{rr'}^{\downarrow\uparrow}\rho_{r'r''}^{\uparrow\uparrow} 
-\rho_{rr'}^{\downarrow\uparrow} h_{r'r''}^{\uparrow\uparrow}
+ h_{rr'}^{\downarrow\downarrow}\rho_{r'r''}^{\downarrow\uparrow} 
-\rho_{rr'}^{\downarrow\downarrow} h_{r'r''}^{\downarrow\uparrow}),
\nonumber\\
&&i\hbar\dot{\rho}_{rr''}^{\downarrow\downarrow} =
\int\!d^3r'(
 h_{rr'}^{\downarrow\uparrow}\rho_{r'r''}^{\uparrow\downarrow} 
-\rho_{rr'}^{\downarrow\uparrow} h_{r'r''}^{\uparrow\downarrow}
+ h_{rr'}^{\downarrow\downarrow}\rho_{r'r''}^{\downarrow\downarrow} 
-\rho_{rr'}^{\downarrow\downarrow} h_{r'r''}^{\downarrow\downarrow}
-\Delta_{rr'}^{\downarrow\uparrow}{\kappa^{\dagger}}_{r'r''}^{\uparrow\downarrow}
+\kappa_{rr'}^{\downarrow\uparrow}{\Delta^{\dagger}}_{r'r''}^{\uparrow\downarrow}),
\nonumber\\
&&i\hbar\dot{\kappa}_{rr''}^{\uparrow\downarrow} = -\Delta_{rr''}^{\uparrow\downarrow}
+\int\!d^3r'\left(
 h_{rr'}^{\uparrow\uparrow}\kappa_{r'r''}^{\uparrow\downarrow} 
+\kappa_{rr'}^{\uparrow\downarrow} {h^*}_{r'r''}^{\downarrow\downarrow}
+\Delta_{rr'}^{\uparrow\downarrow}{\rho^*}_{r'r''}^{\downarrow\downarrow} 
+\rho_{rr'}^{\uparrow\uparrow}\Delta_{r'r''}^{\uparrow\downarrow}
\right),
\nonumber\\
&&i\hbar\dot{\kappa}_{rr''}^{\downarrow\uparrow} = -\Delta_{rr''}^{\downarrow\uparrow}
+\int\!d^3r'\left(
 h_{rr'}^{\downarrow\downarrow}\kappa_{r'r''}^{\downarrow\uparrow} 
+\kappa_{rr'}^{\downarrow\uparrow} {h^*}_{r'r''}^{\uparrow\uparrow}
+\Delta_{rr'}^{\downarrow\uparrow}{\rho^*}_{r'r''}^{\uparrow\uparrow} 
+\rho_{rr'}^{\downarrow\downarrow}\Delta_{r'r''}^{\downarrow\uparrow}
\right).
\label{HFsp}
\end{eqnarray}
with the conventional notation 
$$\uparrow \, \mbox{for}\quad s=\frac{1}{2} \quad \mbox{and}
\quad\downarrow \, \mbox{for}\quad s=-\frac{1}{2}.$$
This set of equations must be complemented by the complex conjugated equations.
Writing these equations we neglected the diagonal in spin matrix elements
of the anomalous density:
$\kappa_{rr'}^{ss}$ and $\Delta_{rr'}^{ss}$. It was shown in~\cite{BaMoPRC2} 
that such an approximation works very well in the case of monopole pairing 
considered here. 

We will consider the Wigner transform \cite{Ring} of equations (\ref{HFsp})
(see \cite{Malov, Malov1} for mathematical details). 
So, instead of four (in $ss'$) matrix elements of
the density matrix $\rho_{rr'}^{ss'}$ and two matrix elements
$\kappa_{rr'}^{\uparrow\downarrow}$, $\kappa_{rr'}^{\downarrow\uparrow}$
we will consider four Wigner functions $f^{ss'}(\br,\bp)$ and two phase space
distributions $\kappa^{ss'}(\br,\bp)$, that is more convenient for WFM method 
(see \cite{BaMoPRC2, BaMoPRC22} for details). From now on, we will not write out the coordinate 
dependence $(\br,\bp)$ of all functions in order to make the formulae 
more transparent. We have
\begin{eqnarray}
      i\hbar\dot f^{\uparrow\uparrow} &=&i\hbar\{h^{\uparrow\uparrow},f^{\uparrow\uparrow}\}
+h^{\uparrow\downarrow}f^{\downarrow\uparrow}-f^{\uparrow\downarrow}h^{\downarrow\uparrow}
+\frac{i\hbar}{2}\{h^{\uparrow\downarrow},f^{\downarrow\uparrow}\}
-\frac{i\hbar}{2}\{f^{\uparrow\downarrow},h^{\downarrow\uparrow}\}
\nonumber\\
&-&\frac{\hbar^2}{8}\{\!\{h^{\uparrow\downarrow},f^{\downarrow\uparrow}\}\!\}
+\frac{\hbar^2}{8}\{\!\{f^{\uparrow\downarrow},h^{\downarrow\uparrow}\}\!\} 
+ \kappa\Delta^* - \Delta\kappa^* 
\nonumber\\
&+&\frac{i\hbar}{2}\{\kappa,\Delta^*\}-\frac{i\hbar}{2}\{\Delta,\kappa^*\}
- \frac{\hbar^2}{8}\{\!\{\kappa,\Delta^*\}\!\} + \frac{\hbar^2}{8}\{\!\{\Delta,\kappa^*\}\!\}
+...,
\nonumber\\
      i\hbar\dot f^{\downarrow\downarrow} &=&i\hbar\{h^{\downarrow\downarrow},f^{\downarrow\downarrow}\}
+h^{\downarrow\uparrow}f^{\uparrow\downarrow}-f^{\downarrow\uparrow}h^{\uparrow\downarrow}
+\frac{i\hbar}{2}\{h^{\downarrow\uparrow},f^{\uparrow\downarrow}\}
-\frac{i\hbar}{2}\{f^{\downarrow\uparrow},h^{\uparrow\downarrow}\}
\nonumber\\
&-&\frac{\hbar^2}{8}\{\!\{h^{\downarrow\uparrow},f^{\uparrow\downarrow}\}\!\}
+\frac{\hbar^2}{8}\{\!\{f^{\downarrow\uparrow},h^{\uparrow\downarrow}\}\!\} 
+ \bar\Delta^* \bar\kappa - \bar\kappa^* \bar\Delta
\nonumber\\
&+&\frac{i\hbar}{2}\{\bar\Delta^*,\bar\kappa\}-\frac{i\hbar}{2}\{\bar\kappa^*,\bar\Delta\}
- \frac{\hbar^2}{8}\{\!\{\bar\Delta^*,\bar\kappa\}\!\} + \frac{\hbar^2}{8}\{\!\{\bar\kappa^*,\bar\Delta\}\!\}
+...,
\nonumber\\
      i\hbar\dot f^{\uparrow\downarrow} &=&
f^{\uparrow\downarrow}h^-
+\frac{i\hbar}{2}\{h^+,f^{\uparrow\downarrow}\}
-\frac{\hbar^2}{8}\{\!\{h^-,f^{\uparrow\downarrow}\}\!\}
\nonumber\\
&-&h^{\uparrow\downarrow}f^-
+\frac{i\hbar}{2}\{h^{\uparrow\downarrow},f^+\}
+\frac{\hbar^2}{8}\{\!\{h^{\uparrow\downarrow},f^-\}\!\}+....,
\nonumber\\
      i\hbar\dot f^{\downarrow\uparrow} &=&
-f^{\downarrow\uparrow}h^-
+\frac{i\hbar}{2}\{h^+,f^{\downarrow\uparrow}\}
+\frac{\hbar^2}{8}\{\!\{h^-,f^{\downarrow\uparrow}\}\!\}
\nonumber\\
&+&h^{\downarrow\uparrow}f^-
+\frac{i\hbar}{2}\{h^{\downarrow\uparrow},f^+\}
-\frac{\hbar^2}{8}\{\!\{h^{\downarrow\uparrow},f^-\}\!\}+...,
\nonumber\\
      i\hbar\dot \kappa &=& \kappa\,(h^{\uparrow\uparrow}+\bar h^{\downarrow\downarrow})
  +\frac{i\hbar}{2}\{(h^{\uparrow\uparrow}-\bar h^{\downarrow\downarrow}),\kappa\}    
  -\frac{\hbar^2}{8}\{\!\{(h^{\uparrow\uparrow}+\bar h^{\downarrow\downarrow}),\kappa\}\!\}   
 \nonumber\\ 
 &+&\Delta\,(f^{\uparrow\uparrow}+\bar f^{\downarrow\downarrow})
  +\frac{i\hbar}{2}\{(f^{\uparrow\uparrow}-\bar f^{\downarrow\downarrow}),\Delta\}    
  -\frac{\hbar^2}{8}\{\!\{(f^{\uparrow\uparrow}+\bar f^{\downarrow\downarrow}),\Delta\}\!\}
  - \Delta + ...,
 \nonumber\\     
    i\hbar\dot \kappa^* &=&  -\kappa^*(h^{\uparrow\uparrow}+\bar h^{\downarrow\downarrow}) 
  +\frac{i\hbar}{2}\{(h^{\uparrow\uparrow}-\bar h^{\downarrow\downarrow}),\kappa^*\}    
  +\frac{\hbar^2}{8}\{\!\{(h^{\uparrow\uparrow}+\bar h^{\downarrow\downarrow}),\kappa^*\}\!\}   
  \nonumber\\
  &-& \Delta^*(f^{\uparrow\uparrow}+\bar f^{\downarrow\downarrow}) 
  +\frac{i\hbar}{2}\{(f^{\uparrow\uparrow}-\bar f^{\downarrow\downarrow}),\Delta^*\}    
  +\frac{\hbar^2}{8}\{\!\{(f^{\uparrow\uparrow}+\bar f^{\downarrow\downarrow}),\Delta^*\}\!\}
  + \Delta^* +...,\qquad 
\label{WHF}
\end{eqnarray} 
where the functions $h^{s,s'}(\br,\bp)$, $f^{s,s'}(\br,\bp)$, $\Delta^{s,s'}(\br,\bp)$, and $\kappa^{s,s'}(\br,\bp)$ are the Wigner
transforms of $h^{s,s'}_{r,r'}$, $\rho^{s,s'}_{r,r'}$, $\Delta^{s,s'}_{r,r'}$, and $\kappa^{s,s'}_{r,r'}$,
respectively, $\bar f(\br,\bp)=f(\br,-\bp)$,
 $\{f,g\}$ is the Poisson
bracket of the functions $f(\br,\bp)$ and $g(\br,\bp)$,
$\{\{f,g\}\}$ is their double Poisson bracket,
$f^{\pm}=f^{\uparrow\uparrow} {\pm} f^{\downarrow\downarrow}$
and $h^{\pm}=h^{\uparrow\uparrow} \pm h^{\downarrow\downarrow}$.
The dots stand for terms proportional to higher powers of $\hbar$ -- after 
integration over phase space these terms disappear and we arrive to the set of
exact integral equations.
This set of equations must be complemented by the dynamical equations for 
$\bar f^{\uparrow\uparrow}, \bar f^{\downarrow\downarrow}, \bar f^{\uparrow\downarrow}, 
\bar f^{\downarrow\uparrow},\bar\kappa,\bar\kappa^*$.
They are obtained by the change $\bp \rightarrow -\bp$ in arguments of functions and Poisson brackets. 
So, in reality we deal with the set of twelve equations. We introduced the notation
$\kappa \equiv \kappa^{\uparrow\downarrow}$ and $\Delta \equiv \Delta^{\uparrow\downarrow}$.
Symmetry properties of matrices $\hat\kappa, \hat\Delta$ and the properties of their Wigner transforms allow one to replace the functions 
$\kappa^{\downarrow\uparrow}(\br,\bp)$ and  $\Delta^{\downarrow\uparrow}(\br,\bp)$ by the functions $\bar\kappa^{\uparrow\downarrow}(\br,\bp)$ and  $\bar\Delta^{\uparrow\downarrow}(\br,\bp)$.

 The microscopic Hamiltonian of the model, harmonic oscillator with 
spin orbit potential plus separable quadrupole-quadrupole and 
spin-spin residual interactions is given by
\begin{eqnarray}
\label{Ham}
 H=\sum\limits_{i=1}^A\left[\frac{\hat\bp_i^2}{2m}+\frac{1}{2}m\omega^2\br_i^2
-\eta\hat \bl_i\hat \bS_i\right]+H_{qq}+H_{ss},
\end{eqnarray}
with
\begin{eqnarray}
\label{Hqq}
&& H_{qq}=\!
\sum_{\mu=-2}^{2}(-1)^{\mu}
\left\{\bar{\kappa}
 \sum\limits_i^Z\!\sum\limits_j^N
+\frac{\kappa}{2}
\left[\sum\limits_{i,j(i\neq j)}^{Z}
+\sum\limits_{i,j(i\neq j)}^{N}
\right]
\right\}
q_{2-\mu}(\br_i)q_{2\mu}(\br_j)
,
\\
\label{Hss}
&&H_{ss}=\!
\sum_{\mu=-1}^{1}(-1)^{\mu}
\left\{\bar{\chi}
 \sum\limits_i^Z\!\sum\limits_j^N
+\frac{\chi}{2}
\left[
\sum\limits_{i,j(i\neq j)}^{Z}
+\sum\limits_{i,j(i\neq j)}^{N}
\right]
\right\}
\hat S_{-\mu}(i)\hat S_{\mu}(j)
\,\delta(\br_i-\br_j),
\end{eqnarray}
where  
$\displaystyle q_{2\mu}(\br)=\sqrt{16\pi/5}\,r^2Y_{2\mu}(\theta,\phi)=
\sqrt{6}\{r\otimes r\}_{2\mu},$
$\{r\otimes r\}_{\lambda\mu}=\sum_{\sigma,\nu}
C_{1\sigma,1\nu}^{\lambda\mu}r_{\sigma}r_{\nu},$
$C_{1\sigma,1\nu}^{\lambda\mu}$
is the Clebsch-Gordan coefficient,
cyclic coordinates $r_{-1}, r_0, r_1$ are defined in~\cite{Var},
$N$ and $Z$ are the numbers of neutrons and protons. 
$\hat S_{\mu}$ are spin matrices~\cite{Var}:
$$
\hat S_1=-\frac{\hbar}{\sqrt2}{0\quad 1\choose 0\quad 0},\
\hat S_0=\frac{\hbar}{2}{1\quad\, 0\choose 0\, -\!1},\
\hat S_{-1}=\frac{\hbar}{\sqrt2}{0\quad 0\choose 1\quad 0}.$$
The mean field generated by this Hamiltonian was derived in \cite{BaMoPRC13}.

\end{widetext}

Equations (\ref{WHF}) will be solved in a small amplitude 
approximation by the WFM method. 
Integrating them over phase space with the weights 
$$W =\{r\otimes p\}_{\lambda\mu},\,\{r\otimes r\}_{\lambda\mu},\,
\{p\otimes p\}_{\lambda\mu}, \mbox{ and } 1$$
one gets dynamic equations for 
the following collective variables:
\begin{eqnarray}
&&\L^{\tau\varsigma}_{\lambda\mu}(t)=\int\! d(\bp,\br) \{r\otimes p\}_{\lambda\mu}
\delta f^{\tau\varsigma}(\br,\bp,t),
\nonumber\\
&&\R^{\tau\varsigma}_{\lambda\mu}(t)=\int\! d(\bp,\br) \{r\otimes r\}_{\lambda\mu}
\delta f^{\tau\varsigma}(\br,\bp,t),
\nonumber\\
&&\P^{\tau\varsigma}_{\lambda\mu}(t)=\int\! d(\bp,\br) \{p\otimes p\}_{\lambda\mu}
\delta f^{\tau\varsigma}(\br,\bp,t),
\nonumber\\
&&\F^{\tau\varsigma}(t)=\int\! d(\bp,\br)
\delta f^{\tau\varsigma}(\br,\bp,t),
\nonumber\\
&&\tilde{\L}^{\tau}_{\lambda\mu}(t)=\int\! d(\bp,\br) \{r\otimes p\}_{\lambda\mu}
\delta \kappa^{\tau i}(\br,\bp,t),
\nonumber\\&&
\tilde{\R}^{\tau}_{\lambda\mu}(t)=\int\! d(\bp,\br) \{r\otimes r\}_{\lambda\mu}
\delta \kappa^{\tau i}(\br,\bp,t),
\nonumber\\
&&\tilde{\P}^{\tau}_{\lambda\mu}(t)=\int\! d(\bp,\br) \{p\otimes p\}_{\lambda\mu}
\delta \kappa^{\tau i}(\br,\bp,t),
\label{Varis}
\end{eqnarray}
 where $\delta f^{\tau\varsigma}$ and $\delta \kappa^{\tau i}$ are variations of $f^{\tau\varsigma}$ and $\kappa^{\tau i}$, 
$\kappa^{\tau i}=\frac{1}{2i}(\kappa^{\tau}-\kappa^{\tau *})$,
$\tau$  denotes  the isospin index,
$\varsigma\!=+,\,-,\,\uparrow\downarrow,\,\downarrow\uparrow$ and
 $\int\! d(\bp,\br)\equiv
(2\pi\hbar)^{-3}\int\! d\br\int\! d\bp$.

The integration yields the sets of coupled (due to neutron-proton interaction)
equations for neutron and proton variables. 
The found equations are nonliner due to quadrupole-quadrupole and spin-spin
interactions. The small amplitude approximation 
allows one to linearize the equations. It is convenient to rewrite them
in terms of isoscalar and isovector variables
$\R_{\lambda\mu}=\R_{\lambda\mu}^{\rm n}+\R_{\lambda\mu}^{\rm p},\quad
\bar \R_{\lambda\mu}=\R_{\lambda\mu}^{\rm n}-\R_{\lambda\mu}^{\rm p}\, ,$
and so on. We also define isovector and isoscalar strength constants
$\kappa_1=\frac{1}{2}(\kappa-\bar\kappa)$ and
$\kappa_0=\frac{1}{2}(\kappa+\bar\kappa)$ connected by the relation
$\kappa_1=\alpha\kappa_0$ with $\alpha=-2$~\cite{BaSc}.
The sets of coupled equations for isovector and isoscalar variables 
are written out in the Appendix~\ref{AppA}.

\section{Results of calculations}\label{III}

The procedure of calculations and parameters of the WFM method are mostly the 
same as in our previous papers \cite{BaMoPRC18,BaMoPRC22}. 
The calculations are performed for the nucleus $^{164}$Dy. 
The solution of 
equations (\ref{iv}, \ref{is}) for $\mu=1$ are presented in the 
Table~\ref{tab1}, where the  energies of $1^+$ levels with their magnetic dipole and electric quadrupole strengths (see~\cite{BaMoPRC22} 
and Appendix~\ref{AppB}) are shown.
Left panel -- the solutions of decoupled equations, right -- isoscalar-isovector coupling is taken into account.

\begin{table}[h!]
\caption{The results of WFM calculations for $^{164}$Dy: energies $E$ (MeV), magnetic dipole $B(M1)$ ($\mu_N^2$) 
and electric quadrupole $B(E2)$ (W.u.) strengths of $1^+$ excitations. 
The marks IS -- isoscalar and IV -- isovector are valid only
for the decoupled case.}\label{tab1}
\begin{ruledtabular}\begin{tabular}{lccccccc}
         \multicolumn{4}{c}{Decoupled equations}  &                       
        &\multicolumn{3}{c}{Coupled equations  }\\  
 \cline{1-4}   \cline{6-8}
     & $E$   & $B(M1)$    & $B(E2)$  & &
       $E$   & $B(M1)$    & $B(E2)$  \\    
 \cline{2-4}   \cline{6-8} 
  IS    &       1.29  & 0.02 & 43.60 &  &  1.47 & 0.17 & 25.44  \\
  IV    &       2.44  & 1.56 &  0.29 &  &  2.20 & 1.76 &  3.30  \\
  IS    &       2.62  & 0.06 &  2.29 &  &  2.87 & 2.24 &  0.34  \\ 
  IV    &       3.35  & 1.08 &  1.29 &  &  3.59 & 1.56 &  4.37  \\
  IS    &      10.94  & 0.00 & 44.36 &  & 10.92 & 0.04 & 50.37  \\ 
  IS    &      14.04  & 0.00 &  2.23 &  & 13.10 & 0.00 &  2.85  \\
  IV    &      14.60  & 0.05 &  0.39 &  & 15.42 & 0.07 &  0.57  \\
  IS    &      15.88  & 0.00 &  0.45 &  & 15.55 & 0.00 &  1.12  \\ 
  IV    &      16.46  & 0.06 &  0.28 &  & 16.78 & 0.06 &  0.53  \\ 
  IS    &      17.69  & 0.00 &  0.36 &  & 17.69 & 0.01 &  0.68  \\
  IS    &      17.90  & 0.00 &  0.41 &  & 17.91 & 0.00 &  0.53  \\ 
  IV    &      18.22  & 0.14 &  1.49 &  & 18.22 & 0.13 &  0.89  \\ 
  IV    &      19.32  & 0.07 &  0.78 &  & 19.32 & 0.08 &  0.61  \\ 
  IV    &      21.29  & 2.00 & 25.26 &  & 21.26 & 2.03 & 21.60  \\            
\end{tabular}\end{ruledtabular}
\end{table} 
Among the high-lying states $\mu=1$ branches of isoscalar (at the energy of $10.92$~MeV) and isovector ($E=21.26$~MeV) 
GQR are distinguished by  large $B(E2)$ values. 
The rest of high-lying states have quite small excitation probabilities and we omit them from further discussion.
Low-lying magnetic states have already been analyzed in~\cite{BaMoPRC22}.
Here the focus will be on the study of the nature of the lowest electrical state.

\begin{figure*}
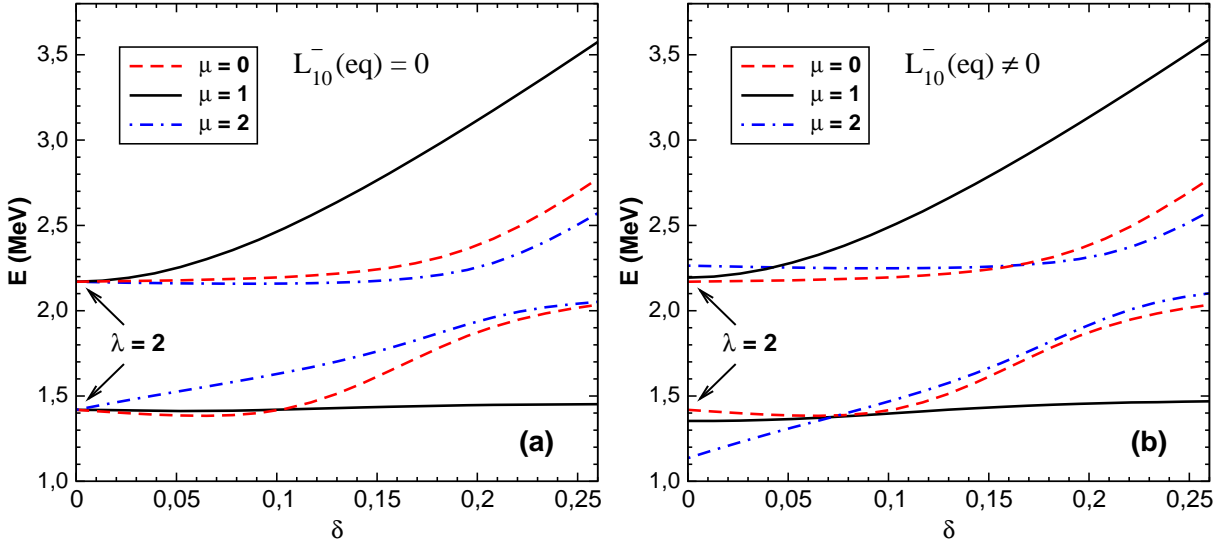

\centering
\includegraphics[width=\columnwidth]{fig1a.eps}\hspace*{2mm}\includegraphics[width=\columnwidth]{fig1b.eps} 
\caption{
Energies $E$ of the low-lying $\mu=0,1,2$ levels as a function of deformation $\delta$.
(a) The calculations were performed without $L^{-}_{10}$(eq).
(b) Splitting of the $2^+$ states into three branches 
in a spherical nucleus as a result of accounting for nuclear antiferromagnetism:
$L^{-}_{10}(\eq)=L^{\uu}_{10}(\eq)-L^{\dd}_{10}(\eq)$.}\label{fig1x}
\end{figure*}

First of all, we note that this state is not a spurious mode. The WFM method ensures the conservation of the total angular 
momentum~\cite{BaMo,BaMoPRC13} (see Appendix~\ref{AppC}) and, as a result, does not produce the spurious  state.
In the absence of coupling, the lowest level with an energy of 1.29~MeV is an isoscalar state of an irrotational nature 
with large $B(E2)$ value and near to zero $B(M1)$ strength.
The isoscalar-isovector coupling strongly affects this state, reducing its $B(E2)$ value by almost half
(from 43~W.u. up to 25~W.u.). However, it retains its electrical nature, 
and its energy changes only from 1.29~MeV to 1.47~MeV (see Table~\ref{tab1}).

The nature of this state can be understood 
after solving dynamical equations for irreducible tensors~(\ref{Varis})
with $\mu=2$ and $\mu=0$ (see Appendix~\ref{AppA}), 
and studying the deformation dependence of the found low-lying levels. 
The results of required calculations for $^{164}$Dy are shown in Fig.~\ref{fig1x}(a).
They demonstrate in an obvious way that the lowest  $1.47$~MeV $1^+$ state 
is just one of three ($K^{\pi}=0^+, 1^+, 2^+$) branches of $J^{\pi}=2^+$ 
state, which can exist in a spherical nucleus (and which splits due to 
deformation into three branches with projections $\mu=0, \pm1, \pm2$).

In a spherical nucleus
the equations for different $\lambda$ are decoupled.
The equations describing the dynamics of $\lambda=2$ variables, in addition to the state that is the subject of our interest, give one more low-lying solution, 
which is a quadrupole vibrational ``ground'' 
of the orbital scissors (see Fig.~\ref{fig1x}(a)). 
As shown in~\cite{BaMoPRC22}, the scissors mode is a mixture of rotational and irrotational flows (see also section~\ref{IV} Currents).
In a spherical nucleus, rotational oscillations are not excited~\cite{Balb22}.
However, due to the coupled dynamics of the conventional scissors mode and isovector GQR~\cite{BaSc, Ann}, 
the vibrational $2^+$ state is excited at the energy of 2.17~MeV with a small value of \mbox{$B(E2)=0.63$~W.u.}. 
As can be seen from the Fig.~\ref{fig1x}(a), 
in the deformed nucleus this level also splits into three branches with different $K^\pi$.

It is necessary to say that there is one more item that must be taken into account in the calculations.
The point is that the results used to draw Fig.~\ref{fig1x}(a) are approximate ones.
They are obtained by neglecting the equilibrium value of 
the collective variable $L^{\tau-}_{10}$(eq), 
which is responsible for the phenomenon of ``hidden angular momenta''~\cite{BaMoPRC2}. 
By definition, $L^{\tau-}_{10}=L^{\tau\uu}_{10}-L^{\tau\dd}_{10}$, where
$L^{\tau\uu}_{10}$ and $L^{\tau\dd}_{10}$ are the average values of the
$z$-component of the orbital angular momentum of all nucleons with the spin projections $\uparrow$ and $\downarrow$ respectively. 
It was shown in~\cite{BaMoPRC2}, that $L^{\tau\uu}_{10}\mbox{\rm(eq)}=-L^{\tau\dd}_{10}$(eq) in the equilibrium state.
So, the ground-state nucleus consists of two equal parts
having nonzero angular momenta with opposite directions, which compensate each other resulting in 
the zero total angular momentum ($L^{\tau+}_{10}\mbox{\rm(eq)}=L^{\tau\uu}_{10}\mbox{\rm(eq)}+L^{\tau\dd}_{10}\mbox{\rm(eq)}=0$), 
whereas $L^{\tau-}_{10}\mbox{\rm(eq)}\neq 0$ even in the spherical limit.
This phenomenon is 
quite similar to the phenomenon of the antiferromagnetism~\cite{Gurevich}, 
so we will use in the following namely this 
term.

The simplest model of an antiferromagnet is one in which it is represented as a set of two equivalent interpenetrating magnetic sublattices 
characterized by the same value but antiparallel oriented magnetization densities~\cite{Akhiezer}.
In the ground state, the magnetic moments of the sublattices are compensated. 
The action of an external magnetic field leads to the appearance of a macroscopic magnetic moment of the antiferromagnet.
The manifestation of antiferromagnetism in our calculations is a consequence of the fact that the WFM method deals with a dynamic 
mean field~\cite{BaMoPRC13,Balb22}. The resulting splitting is the reaction of the mean field of the nucleus to an external perturbation.
The results of calculations, including $L^{\tau-}_{10}$(eq)), are 
shown~in~Fig.~\ref{fig1x}(b).
It is seen that taking into account the antiferromagnetic properties of 
nuclei leads to the splitting of $2^+$ states already at zero deformation. 
It is interesting to note that the energy of the lowest $\mu=1$ state weakly depends on the deformation (see~Figs.~\ref{fig1x}(a),~\ref{fig1x}(b)),
while the $E2$ strength decreases from $25.44$ W.u. to $15.56$ W.u. in the spherical limit.

\begin{figure}
\centering
\includegraphics[width=\columnwidth]{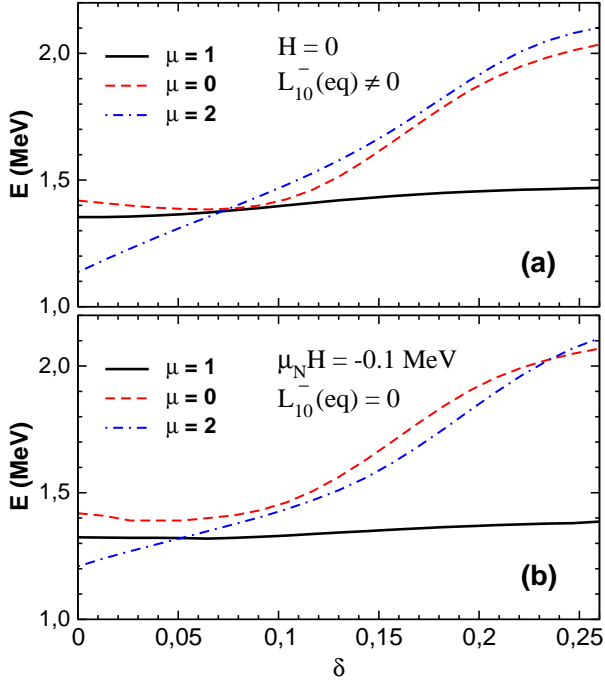}
\caption{The $\mu=0,1,2$ energy branches of nucleus with $N$ and $Z$ corresponding to $^{164}$Dy vs. deformation $\delta$.
The results of calculations with $L_{10}^-\neq 0, \H=0$ (a) and 
$L_{10}^-=0, \H\neq 0$ (b) (see also columns I and III in Table~\ref{TMagnet}).
}
\label{fig2}\end{figure}

To verify the relevance of the analogy with antiferromagnetism, 
we compare the effect of $L^{\tau-}_{10}$(eq)
with the influence of an external magnetic field~\cite{Gurevich, Davydov}. 
In the absence of an external magnetic field 
the states with different projections of the nuclear magnetic moment onto the 
quantization axis are degenerate. 
When an external magnetic field $\H$ is applied, the magnetic moment acquires
the potential energy in this field (Zeeman energy) and, as a result, the 
degeneracy of states with different projections is removed.
According to~\cite{Davydov}, the action of a 
uniform magnetic field  $\H$ can be described in a linear approximation
by adding to the Hamiltonian the term
$\M=-\hat{\mu}\cdot\hat{\H}$,
where 
\begin{equation}
\hat{\mu}=\frac{e}{2mc} \sum\limits_{\tau={\rm p,n}}\left(g_l^\tau\hat{\bf l}^\tau+g_s^\tau\hat{\bf s}^\tau\right)
\end{equation} 
is the operator of the magnetic moment of the nucleus, 
$\hat{\bf l}=[\hat{\br}\times\hat{\bp}]$ -- the orbital angular momentum, 
$\hat{\bp}=-i\hbar\bf{\nabla}$, 
$\hat{\bf s}=\frac{\hbar}{2}\hat{\bf\sigma}$ -- the spin operator, 
$\hat{\sigma}$ -- Pauli matrix, $g_l$ and $g_s$ are orbital and spin 
$g$-factors (see Appendix~\ref{AppB}).

\begin{table}[t!]
\caption{The calculated splitting of $2^+$ state which can exist in spherical 
nucleus with N and Z corresponding to $^{164}$Dy:\\
I -- $L_{10}^-(\eq)=22i\hbar$, $\H=0$,\\  
II -- $L_{10}^-(\eq)=0$, $\mu_N\H=-0.13$ MeV,\\
III -- $L_{10}^-(\eq)=0$, $\mu_N\H=-0.1$ MeV,\\  
IV -- $L_{10}^-(\eq)=0$, $\mu_N\H=-0.05$ MeV.}
\begin{ruledtabular}\begin{tabular}{ccccc}
 $\mu$  & \multicolumn{4}{c}{$E$, MeV}       \\    
 \hline
        & I & II & III & IV\\
 \cline{2-5}
  0 &  1.42 &  1.42 &  1.42 &  1.42  \\
  1 &  1.35 &  1.28 &  1.32 &  1.37  \\
  2 &  1.14 &  1.15 &  1.21 &  1.31  \\
\end{tabular}\end{ruledtabular}
\label{TMagnet}
\end{table}
Let us consider the magnetic field directed along $z$-axis. In this case $\hat{\H}=\H_z\equiv\H$ and
\begin{eqnarray}
 \M^\tau&&=-\frac{e\H}{2mc}(g_l^\tau\hat{l_0}+g_s^\tau\hat{s_0}) \nonumber\\
&&=\frac{e\hbar\H}{2mc}\left(g_l^\tau\sqrt{2}\{\hat{r}\otimes\nabla\}_{10}-\frac{g_s^\tau}{2}\hat{\sigma_0}\right). 
 \nonumber\end{eqnarray}
The Wigner transformation of $\M^\tau$ reads:
\begin{equation}
\M^\tau_W=\frac{e\H}{2mc}\left(g_l^\tau i
\sqrt{2}\{r\otimes p\}_{10}
-g_s^\tau\frac{\hbar}{2}{\sigma_0}\right).  
\nonumber\end{equation}
By definition $h^{\pm}=h^{\uparrow\uparrow}\pm h^{\downarrow\downarrow}$.
Remembering that $\sigma_0={1\quad\, 0\choose \,0\;\;\; -\!1}$~\cite{Var}, one finds
\begin{equation}
 h_W^{\tau+}=i \mu_N\H g_l^\tau\frac{2\sqrt2}{\hbar}\{r\otimes p\}_{10},\quad
 h_W^{\tau-}=-g_s^\tau\,\mu_N\H,
 \nonumber
\end{equation}
where $\mu_N=\frac{e\hbar}{2mc}$ is the nuclear magneton.
The contribution of the magnetic field into dynamical equations for variables
$\L^{\tau\varsigma}_{\lambda\mu}(t)$,
$\R^{\tau\varsigma}_{\lambda\mu}(t)$,
$\P^{\tau\varsigma}_{\lambda\mu}(t)$ 
is written out in Appendix~\ref{AppA}. There are 24 equations for $\mu=2$, 44 equations for
$\mu=1$ and 52 ones for $\mu=0$. 
It is worth to note, that magnetic field contributes into all 24 equations for 
$\mu=2$, into 38 equations for $\mu=1$ and  into 16 equations for $\mu=0$,  whereas 
$L_{10}^-$(eq) enters only in two equations for $\mu=1$ and $\mu=2$, and does not contribute to the equations for $\mu=0$. 
Nevertheless their influence on the splitting of $2^+$ 
state turns out quite close (at the proper choice of the magnetic field strength 
$\H$). At the zero deformation, the energy of splitting 
(Zeeman energy) is $\Delta E\simeq\mu\H\mu_N$ MeV.
The results of calculations with $L_{10}^-\neq 0, \H=0$ and 
$L_{10}^-=0, \H\neq 0$ for the lowest state are compared in the Table~\ref{TMagnet} and Fig.~\ref{fig2}~(a,~b). 
There are no fundamental differences between the effect produced by nuclear antiferromagnetism and the result obtained 
when a uniform magnetic field is turned on.
Remarkably, the nuclear level spacing $\sim 0.1$ MeV gives respective field strength scale $\H=\Delta E/\mu_N\sim$ 1 TT (Teratesla) = $10^{16}$ G (Gauss), 
which is comparable to the range of magnetic strength of magnetars~\cite{Pena}.

What variables are responsible for the generation of this excitation?
First of all, it is the branch of the splitted 2$^+$ state, which can exist in 
spherical nuclei. Therefore, the respective variables should have $\lambda=2$.
Furthermore, the calculations without pair correlations show, that the energy of this 
state goes to zero if to omit the variables 
$\R^{\uparrow\downarrow}_{22}$,
$\R^{\downarrow\uparrow}_{20}$, 
$\P^{\uparrow\downarrow}_{22}$,
$\P^{\downarrow\uparrow}_{20}$ 
together with their dynamical equations (see Appendix~\ref{AppA}). 
According to the spin indexes, these variables describe spin-flip processes.
Similarly, the solution disappears if variables $\R^{+}_{21}$ and $\P^{+}_{21}$  are omitted together with their dynamical equations.
The natural conclusion is that these two sets of variables are the generators of the excitation. 
This fact allows one to suppose the rather interesting
peculiarity of the excitation process of this nuclear mode: the electric
external field $\hat O_{21}=er^2Y_{21}$ (see Appendix~\ref{AppB}) initiates the 
dynamical deformation $\R^{+}_{21}(t)$ together with the Fermi 
surface deformation $\P^{+}_{21}(t)$, which in their turn initiate spin-flip
transitions 
$\R^{\uparrow\downarrow}_{22}(t)$,
$\R^{\downarrow\uparrow}_{20}(t)$, 
$\P^{\uparrow\downarrow}_{22}(t)$,
$\P^{\downarrow\uparrow}_{20}(t)$.
Our calculations show that
this is the only low-lying state in the formation of which spin-flip variables are involved.

\section{Currents}\label{IV}

The analysis of currents of three low-lying magnetic excitations had shown that
every of them represents rather complicate mixture of all three possible 
scissors modes~\cite{BaMoPRC22}. Besides, as Lipparini and Stringari~\cite{LS83} 
and later Balbutsev {\it et al}~\cite{Malov,Ann}  have shown, microscopically there is a strong 
coupling between the scissors mode and the isovector GQR. 
Actually without this coupling the scissors mode  comes at zero energy. 
The result of this coupling is also that the rotation is accompanied by a quadrupole distortion of the shape of the nucleus, see~\cite{BaSc}.

Now it will be interesting  
to analyze the structure of currents
corresponding to the 
electric excitation disposed at the energy $E=1.47$~MeV, just below all scissors modes. The detailed 
description of the procedure of constructing nuclear currents is given in the
paper~\cite{BaMoPRC22}.
\begin{figure}[t!]\centering
\includegraphics[width=.5\columnwidth]{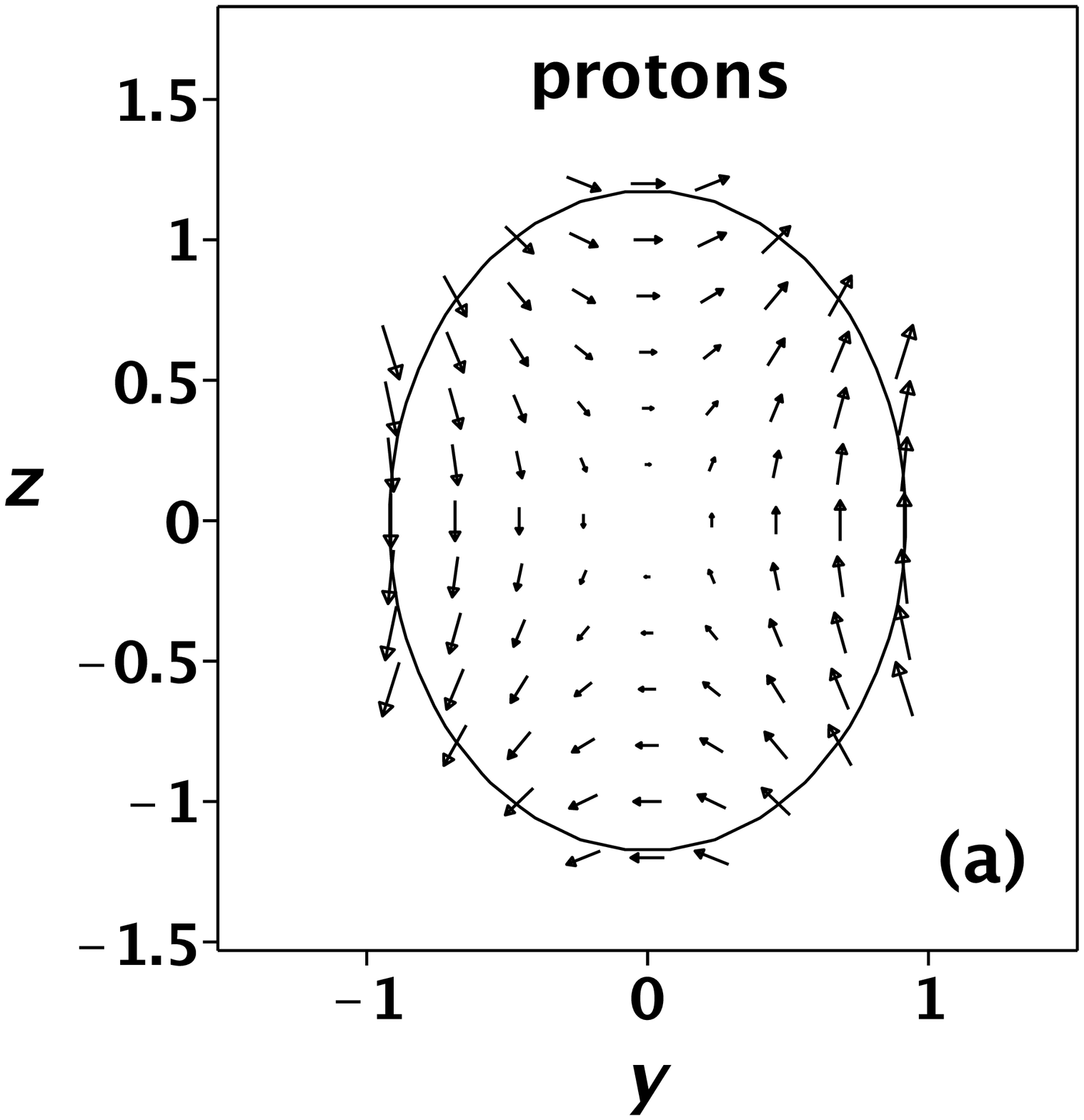}\includegraphics[width=0.5\columnwidth]{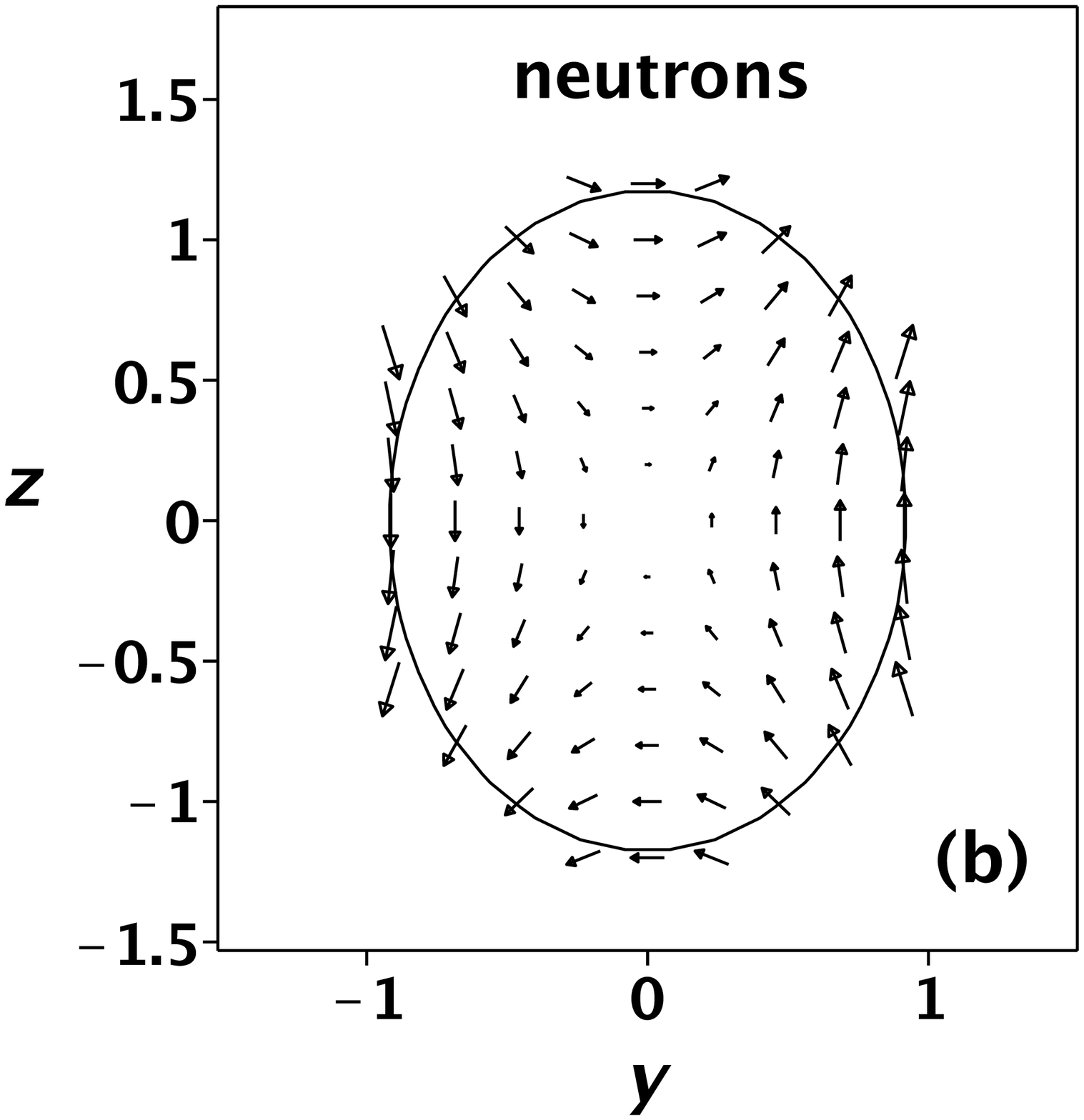}
\caption{The proton and neutron currents in $^{164}$Dy for the lowest electrical level with energy $E=1.29$ MeV. 
The calculations were performed without the isoscalar-isovector coupling. Here and below, a sharp-edge density distribution is used for visualization;
{\sl y $=y/R_0$, z $=z/R_0$}, $R_0$ -- nuclear radius.}
\label{fig1c}\end{figure}
\begin{figure}[b!]\centering
\includegraphics[width=.5\columnwidth]{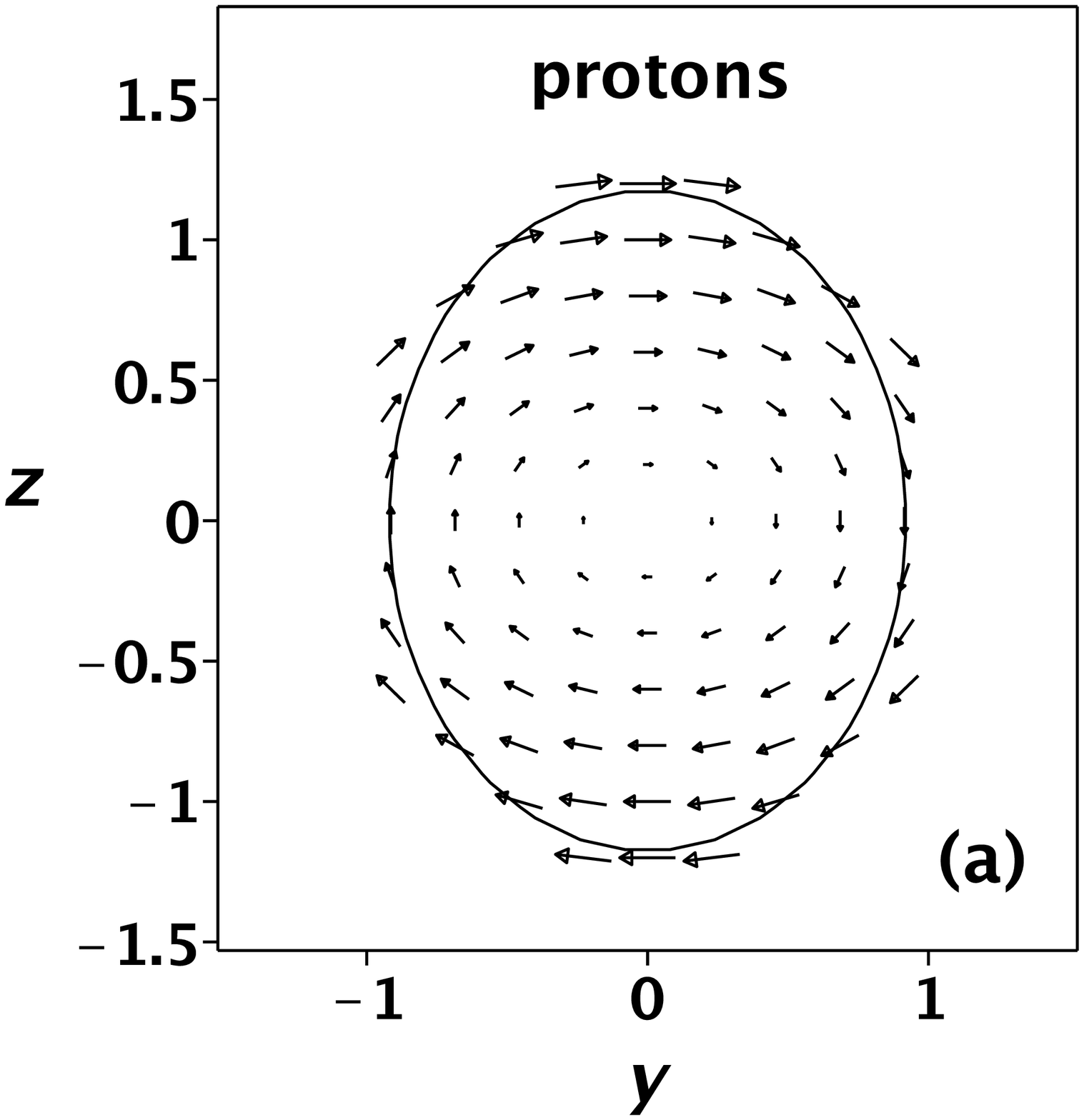}\includegraphics[width=0.5\columnwidth]{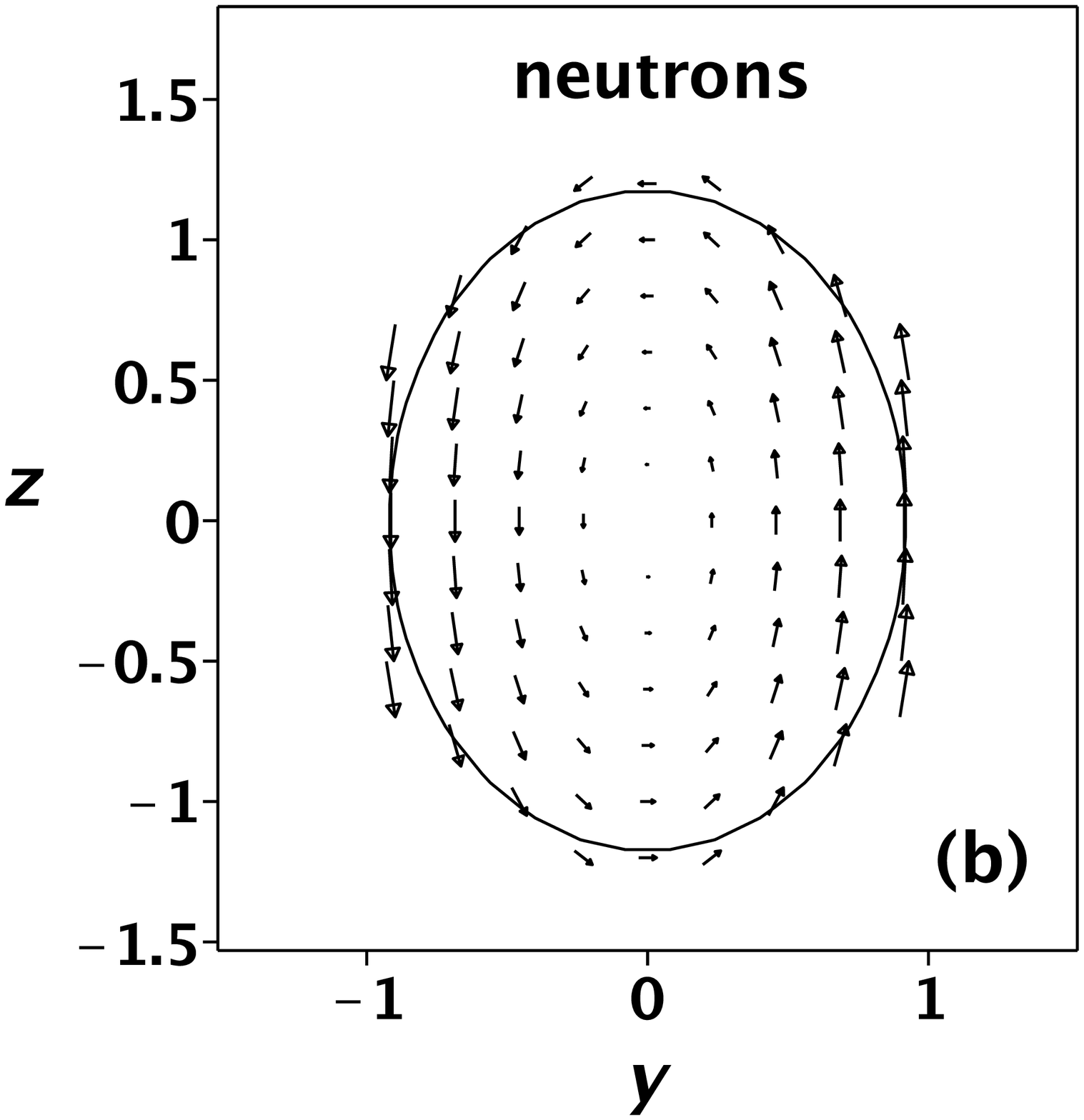}
\includegraphics[width=.5\columnwidth]{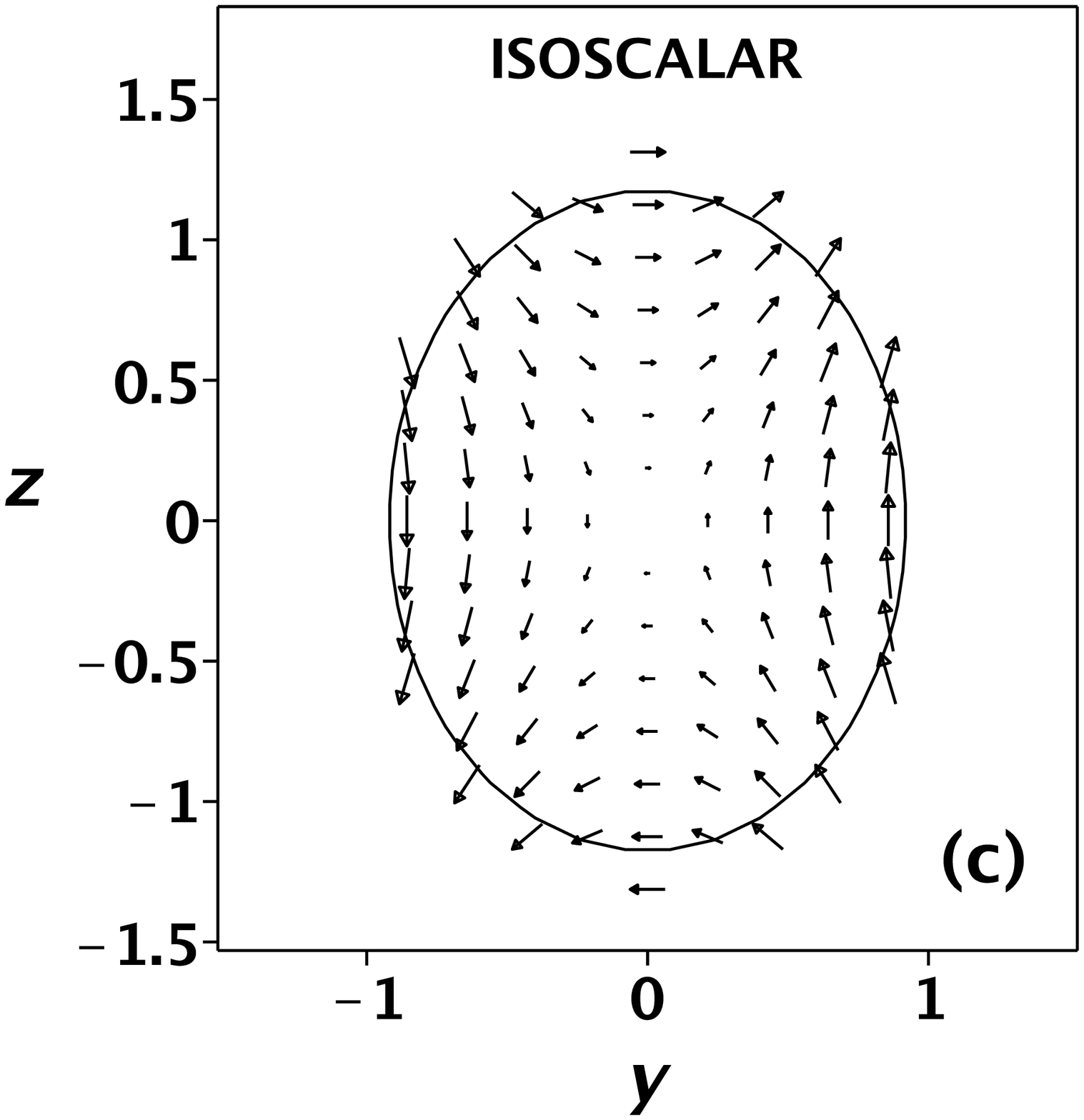}\includegraphics[width=0.5\columnwidth]{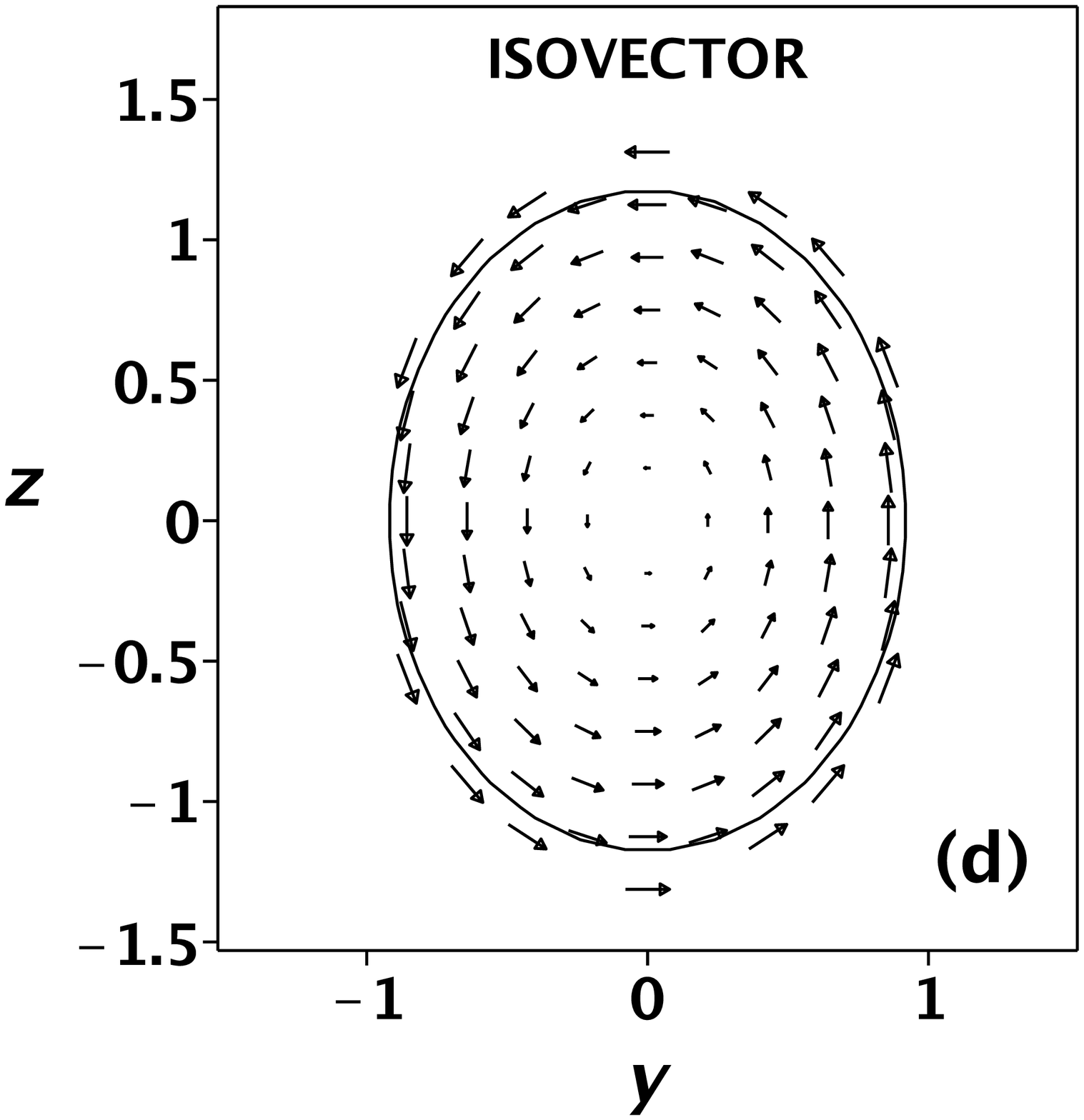}
\caption{The currents in $^{164}$Dy for the lowest electrical level with energy $E=1.47$ MeV. The isoscalar-isovector coupling is taken into account.}
\label{fig1a}\end{figure} 
The resulting spin-scalar current field can be represented as a superposition of rotational and vibrational fields
\begin{eqnarray}
\label{Jc}
 {\bf J}^{\tau}(\br)=n(\br)\left[ a^{\tau}\left[{\bf e}_x\times{\bf r}\right]
 +b^{\tau}\nabla(yz) \right],
\end{eqnarray}
with amplitudes $a^{\tau}$ and $b^{\tau}$  respectively
\begin{eqnarray}
\label{ab}
&&a^{\tau}=3i\,\frac{\L^{+\tau}_{11}\left(1+\delta/3\right)+\delta \L^{+\tau}_{21}}
{Q_{00}\left(1-2/3\delta\right)\left(1+4/3\delta\right)},\nonumber\\
&&b^{\tau}=3i\,\frac{\L^{+\tau}_{21}\left(1+\delta/3\right)+\delta \L^{+\tau}_{11}}
{Q_{00}\left(1-2/3\delta\right)\left(1+4/3\delta\right)},
\end{eqnarray}
where $n(\br)$ -- nuclear density distribution,  $\delta$ -- deformation (see Appendix~\ref{AppA} for parameters).
As can be seen from (\ref{Jc},~\ref{ab}), the rotational contribution is  
determined predominantly by the $\mu=1$ projection of the
collective orbital angular momentum~$\L^{+\tau}_{11}$,
while the vibrational component is determined mainly by the value of the 
variable 
$\L^{+\tau}_{21}$ which represents the velocity of changing of the nuclear 
shape ($\L^{+\tau}_{21}\sim \dot{\R}^{+\tau}_{21}$).
It was shown in \cite{BaMoPRC22} that in the case of three magnetic excitations
the field~(\ref{Jc}) is flat, all motions take place only in one plane, as it 
should be for real scissors. The same is true for the considered excitation.
\begin{figure}[t!]\centering
\includegraphics[width=.5\columnwidth]{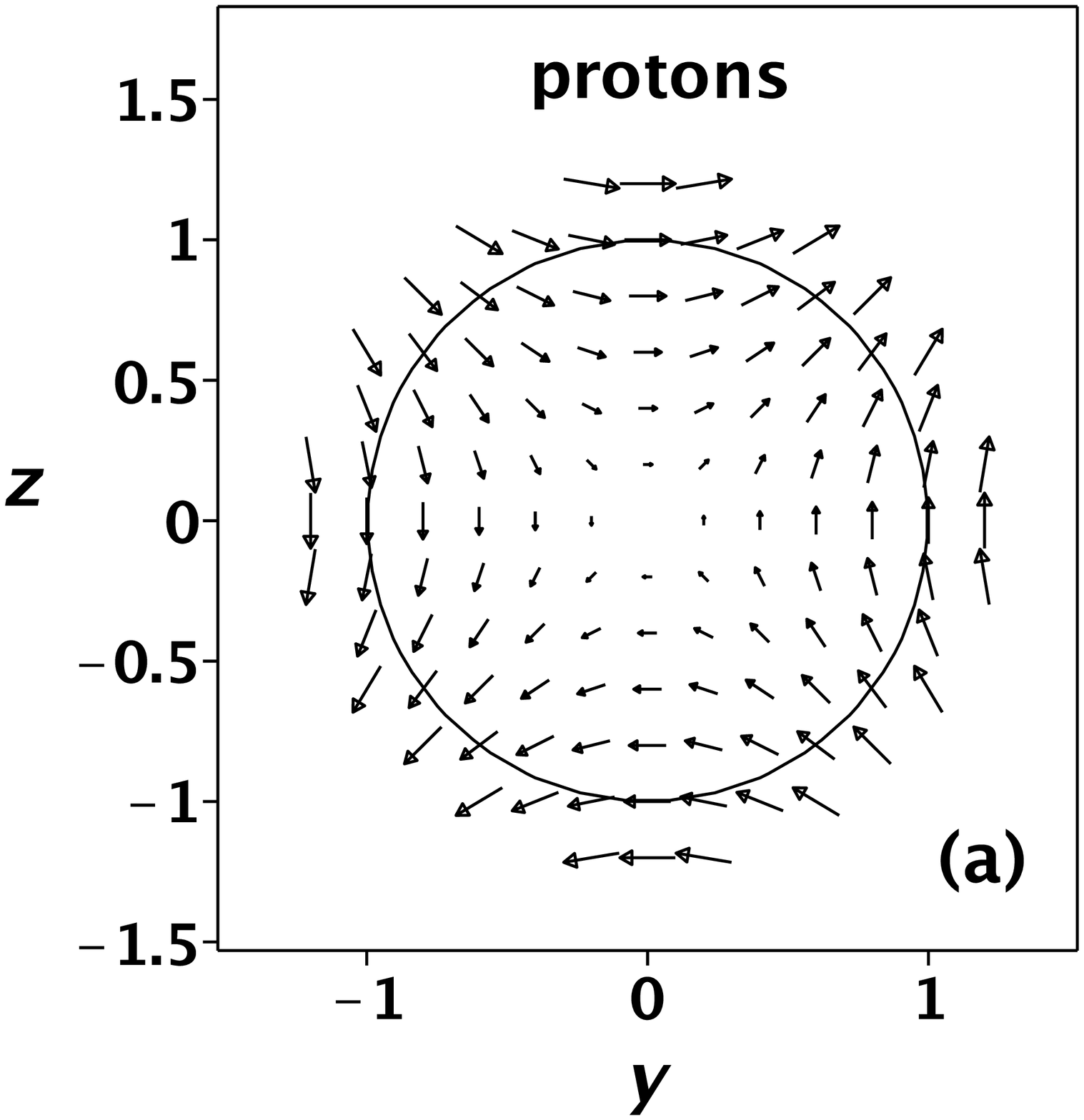}\includegraphics[width=0.5\columnwidth]{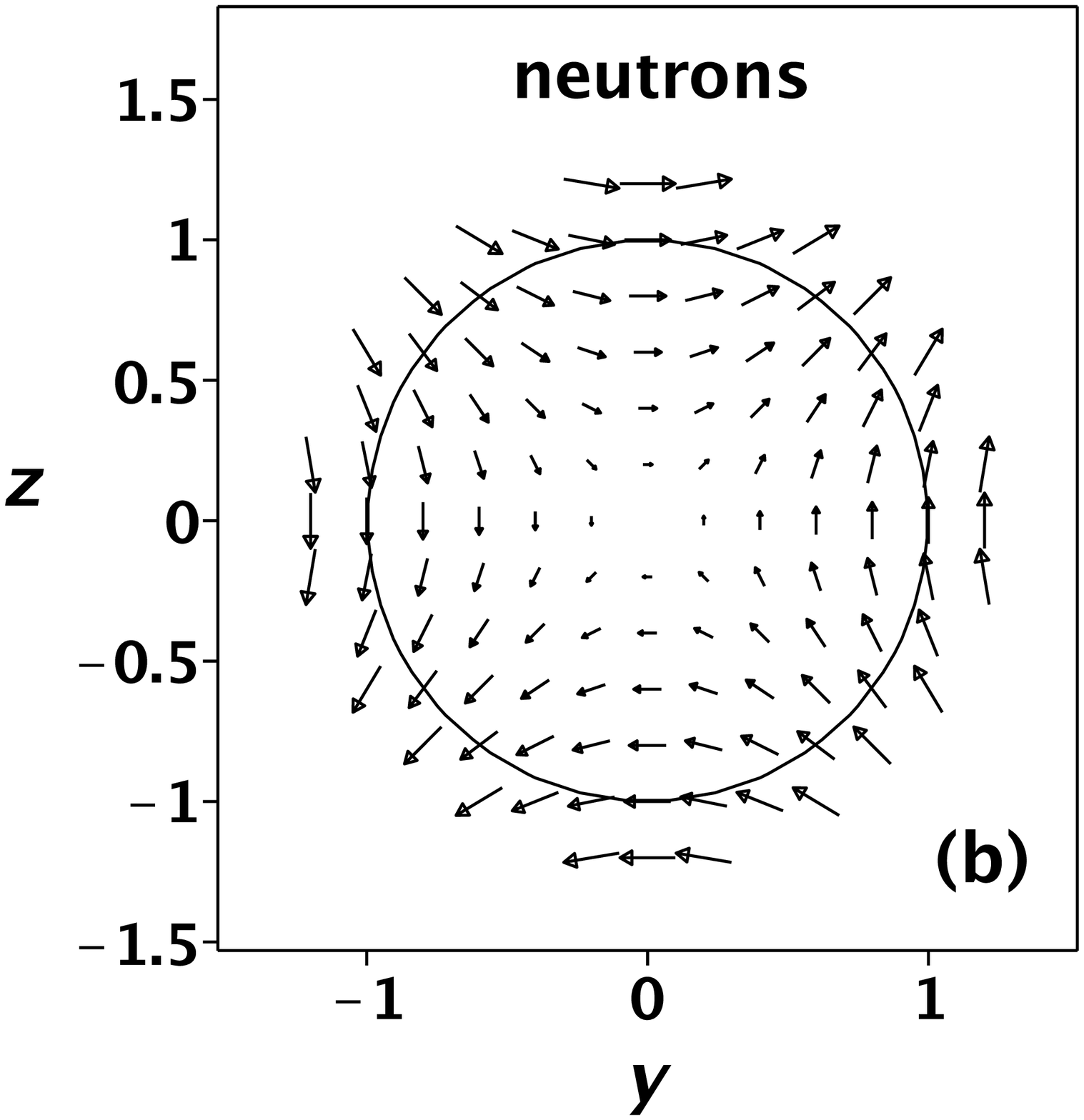}
\includegraphics[width=.5\columnwidth]{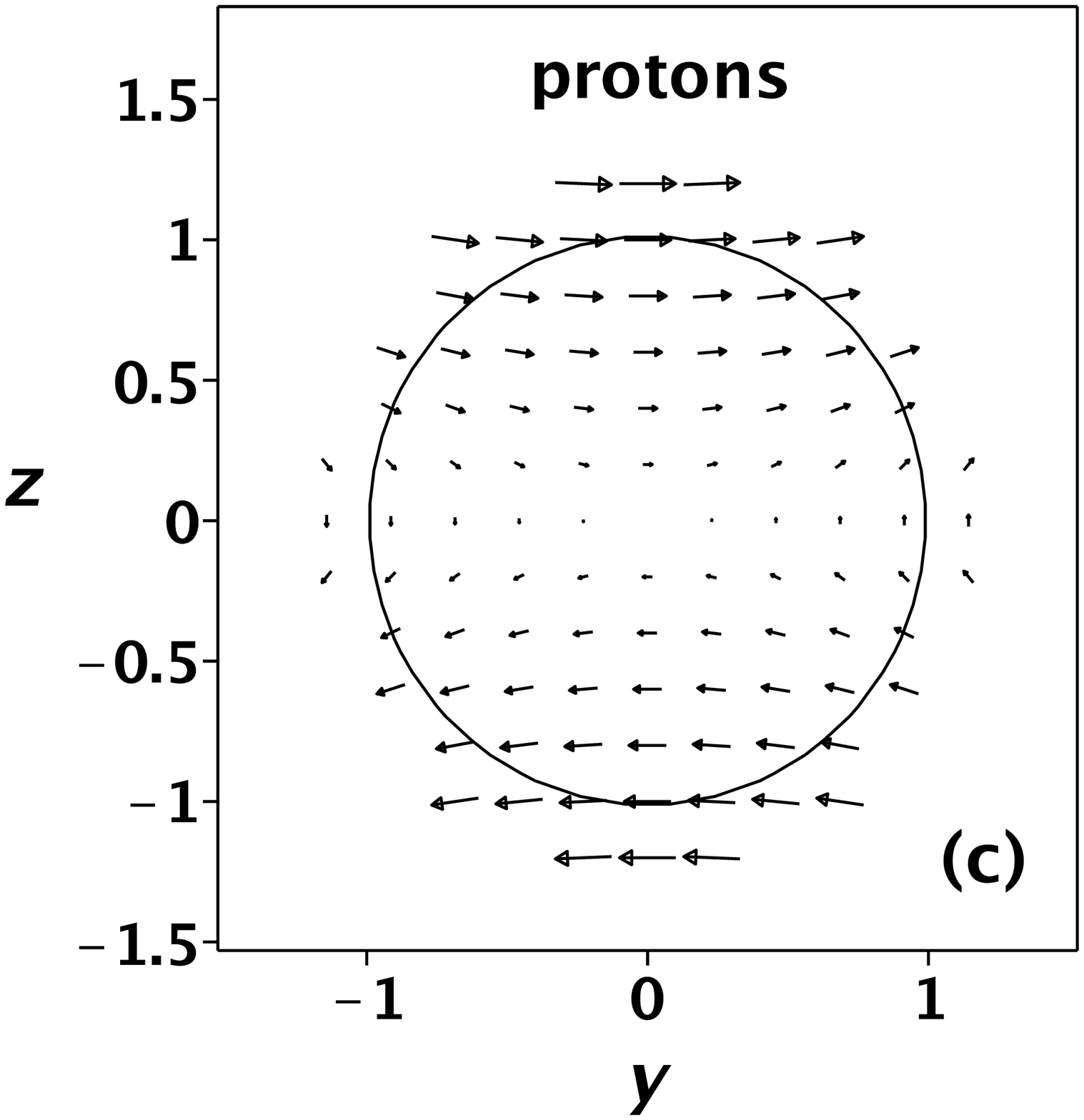}\includegraphics[width=0.5\columnwidth]{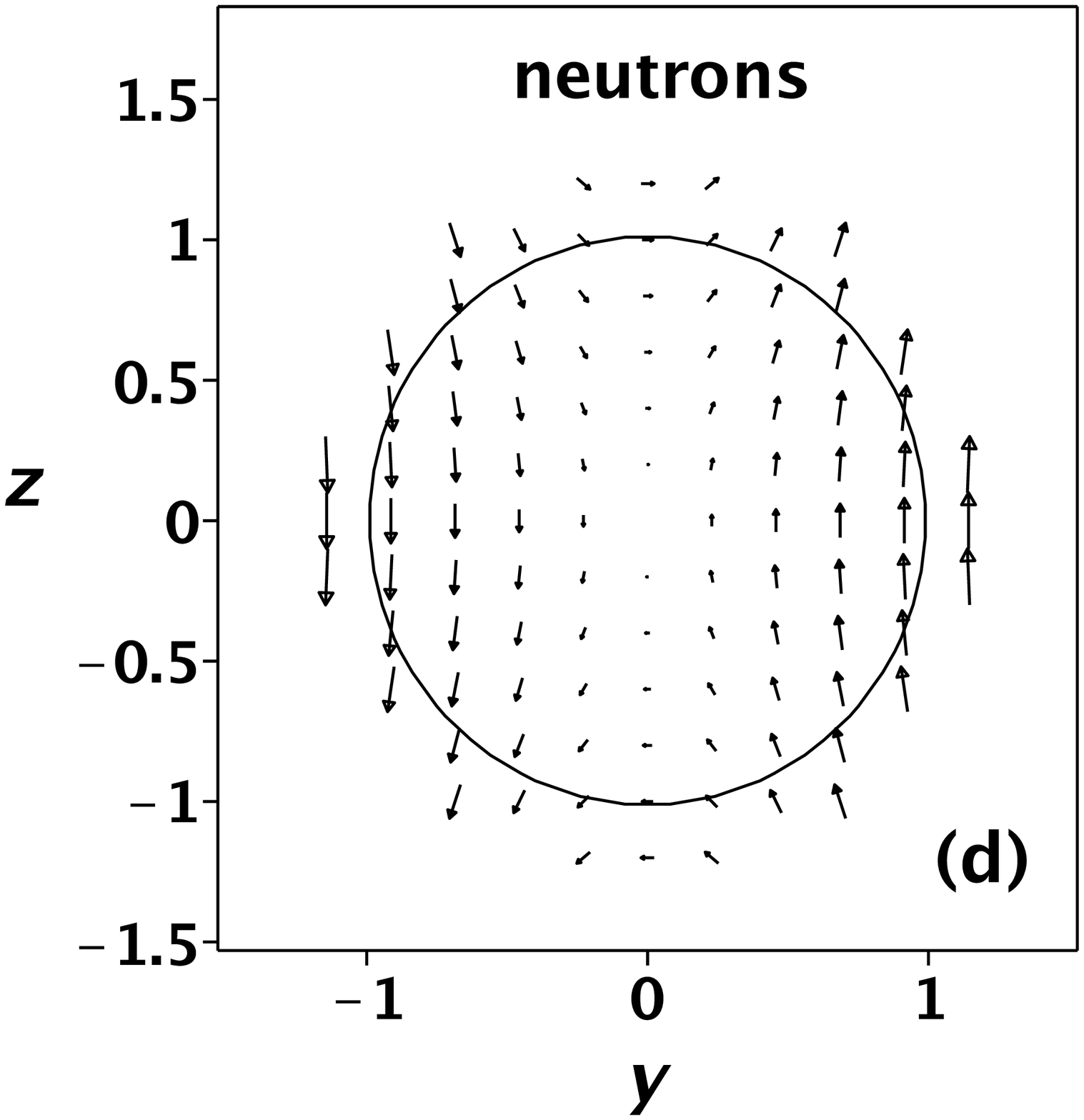}
\includegraphics[width=.5\columnwidth]{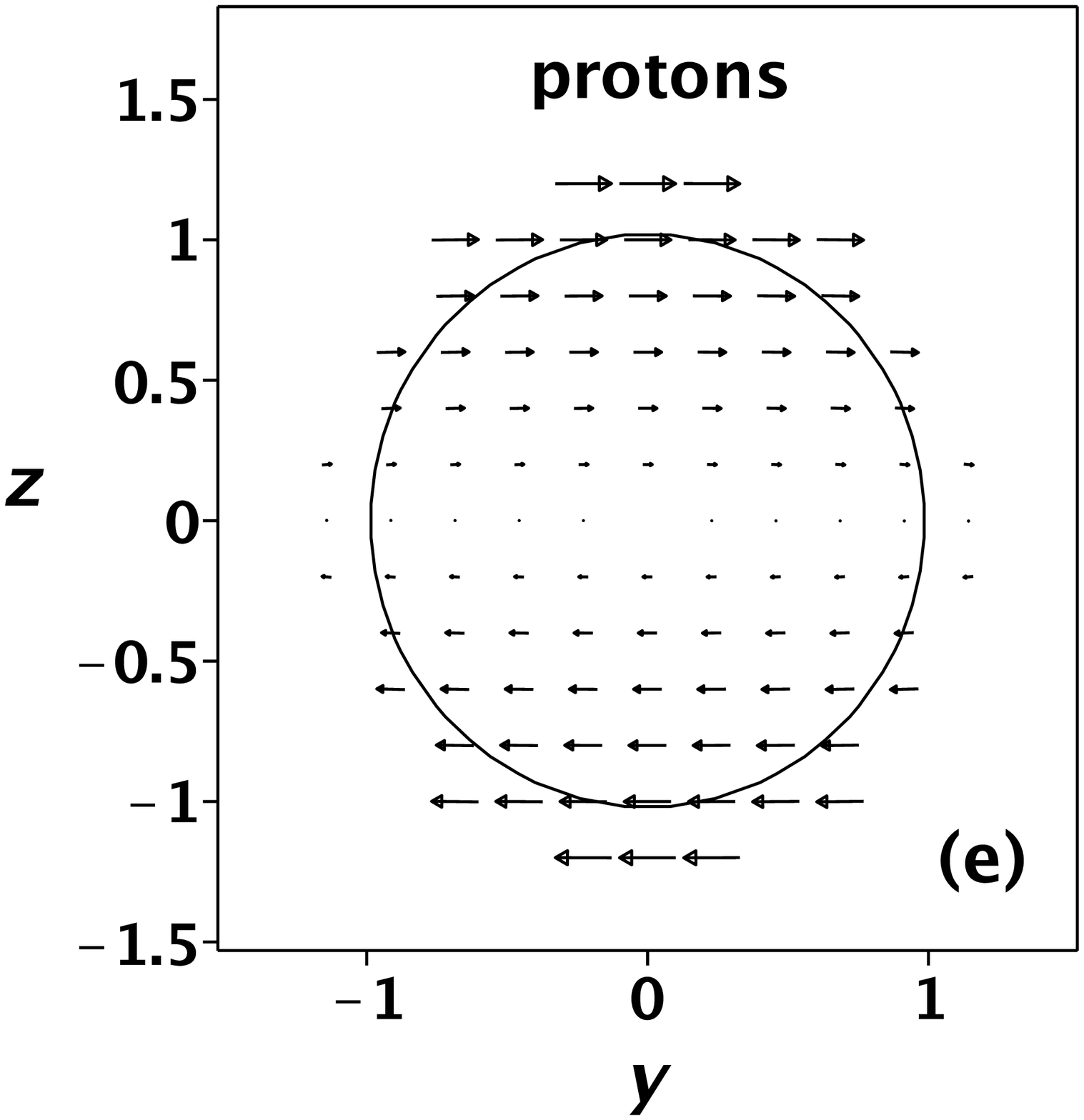}\includegraphics[width=0.5\columnwidth]{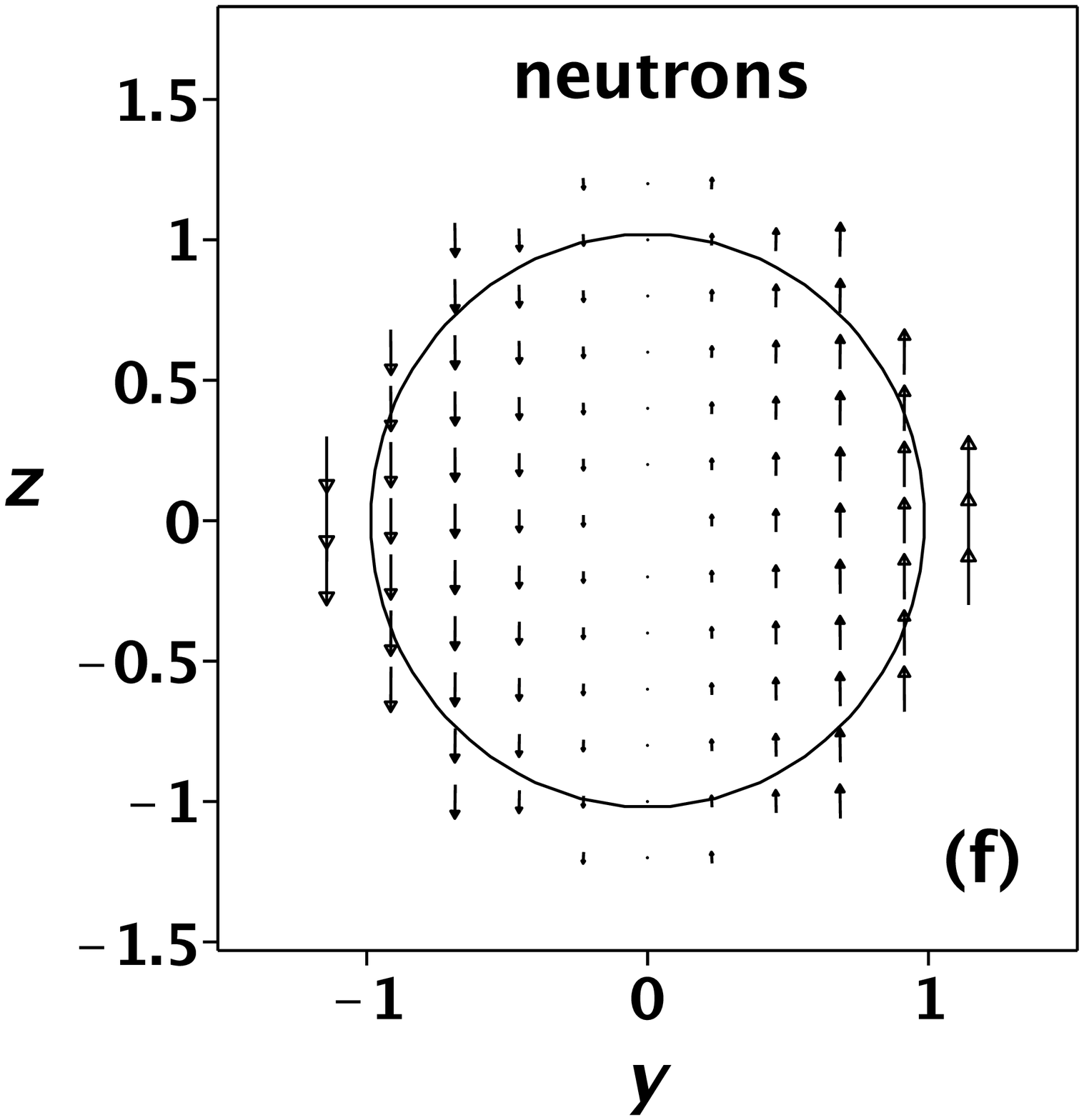}
\includegraphics[width=.5\columnwidth]{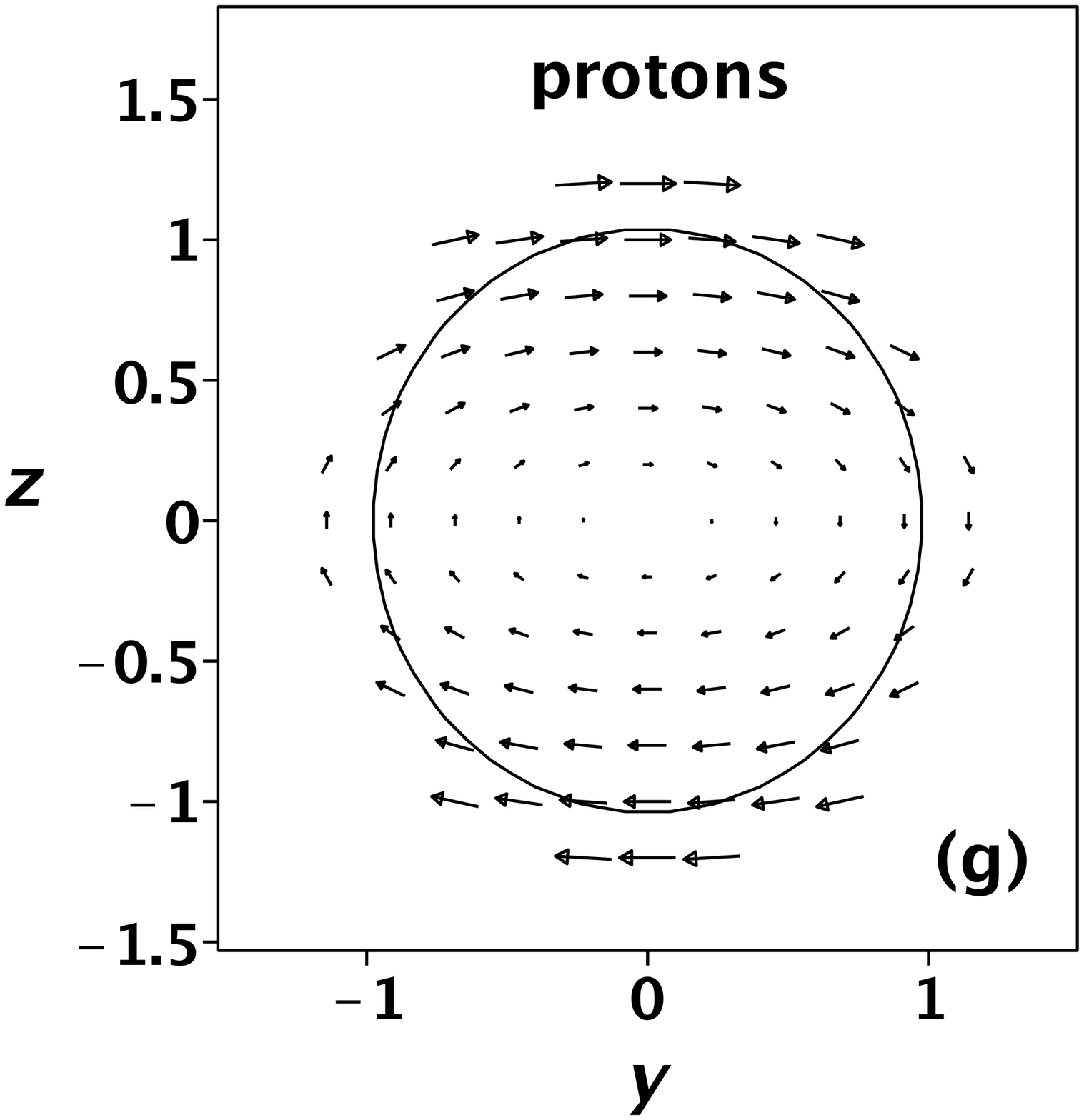}\includegraphics[width=0.5\columnwidth]{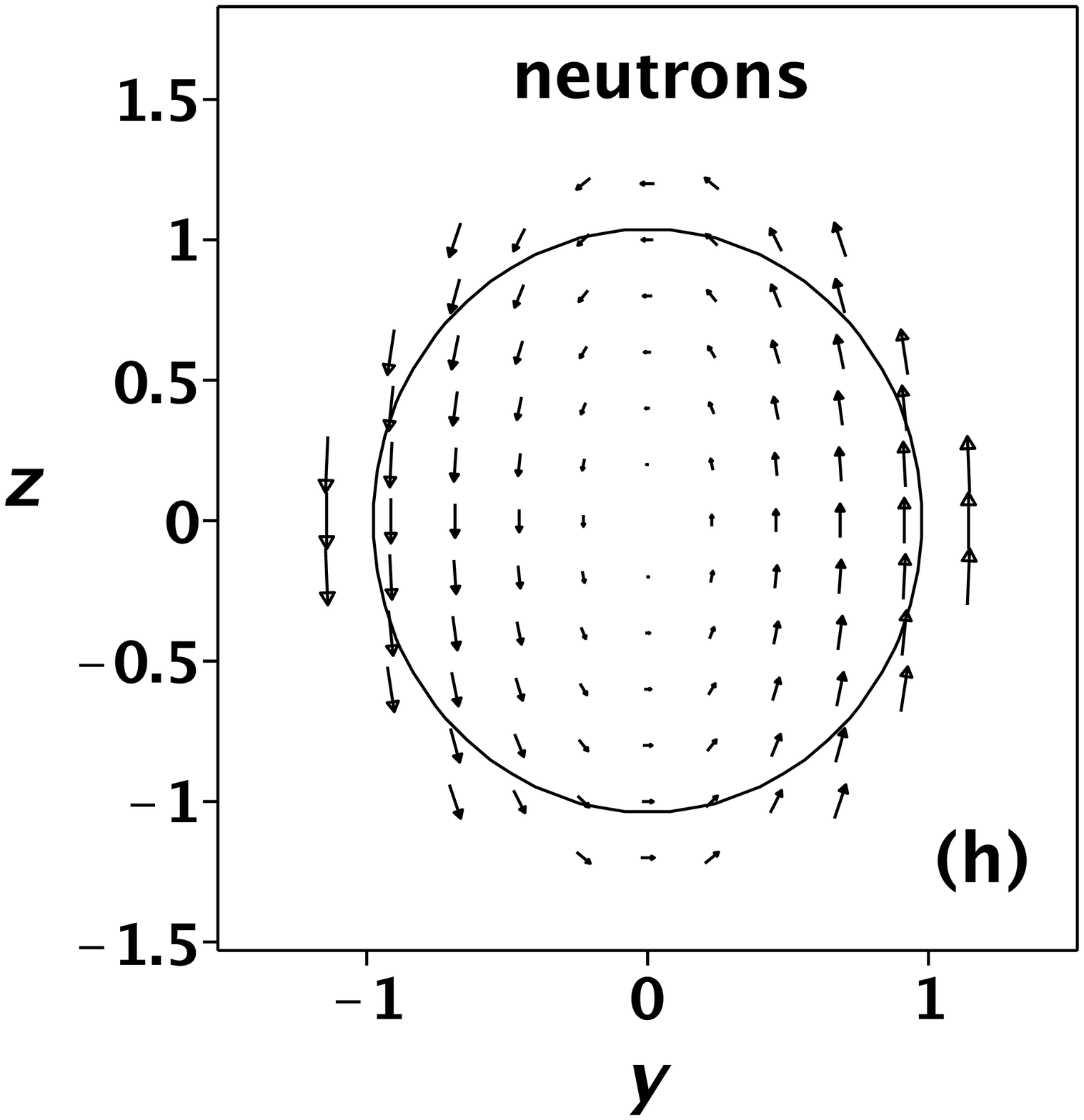}
\caption{The currents in the nucleus with $N$ and $Z$ corresponding to $^{164}$Dy for the lowest electrical level:
(a),(b) -- $\delta=0$, (c),(d) -- $\delta=0.02$, (e),(f) -- $\delta=0.032$, (g),(h) -- $\delta=0.06$.}\label{fig6}
\end{figure}

Calculations performed without the coupling of isovector and isoscalar equations  
demonstrate a clear quadrupole vibrational pattern 
(see Fig.~\ref{fig1c}, where the currents of neutrons and protons are shown
separately). Protons and neutrons move in phase.
This is the expected result, since in such approximation the considered solution is purely isoscalar (see Table~\ref{tab1}).
Taking into account the isovector-isoscalar coupling
greatly changes the pattern of flows.
As can be seen from Fig.~\ref{fig1a}~(a,~b), both currents acquire a rotational character.
The trajectories of 
nucleons are the strongly stretched (prolate) ellipses, their long axes being 
perpendicular one to another. The overall picture is close to a shear displacements.
The sum of proton and neutron currents is shown in Fig.~\ref{fig1a}~(c). 
Surprisingly, the character of the displacements turns out 
almost pure vibrational. This result is confirmed by the rather large $B(E2)$ value (see Table~\ref{tab1}, first line). 
The isovector contribution is shown in Fig.~\ref{fig1a}~(d). It is almost purely rotational, that obviously corresponds to the conventional scissors
(see Table~\ref{tab1}, fourth line).

Let us stress, that we define the isoscalar (isovector) mode as the sum
(difference) of neutron and proton amplitudes, independently of the character
of their relative motion -- in phase or out of phase. For example, it is
impossible to say anything definite about the relative phase of neutron and 
proton currents in Figs.~\ref{fig1a}~(a) and~\ref{fig1a}~(b). Nevertheless their sum and difference
(Figs.~\ref{fig1a}~(c) and~\ref{fig1a}~(d)) produce evident pictures of the isoscalar vibrational and
the isovector rotational motions.

Finally, it is interesting to investigate the behavior of currents in the spherical limit ($\delta=0$). 
The collective rotation is completely absent at zero deformation.
So in this case, as expected, we obtain an undoubtedly quadrupole vibrational picture for considered constituents, 
which is shown in the Fig.~\ref{fig6}~(a,~b). 
As a consequence, this leads to an increase in the value of $B(E2)$ up to 43 W.u. and $B(M1)=0$ $\mu_N^2$.
But even negligibly small deviation from a spherical shape immediately transforms the vibrational motion
of protons and neutrons into the ``shear-rotational'' one.
The evolution of the flow distributions with increasing deformation is tracked down in Fig.~\ref{fig6}~(a,~b,~c,~d,~e,~f,~g,~h). 
Such behavior is a direct effect of the isoscalar-isovector coupling, which enables the admixture of the scissors rotation.
Thus, the analysis shows that in even-even heavy nuclei, the collective motion of protons and neutrons 
associated with low-energy quadrupole vibrational excitations is strongly influenced by the scissors-like rotation. 
Although the summed flow distribution of nucleons retains its vibrational character, the isovector contribution has an appreciable effect.

We are only considering convection currents here. 
As discussed in the previous section, spin-flip processes are essential for the formation of the state under consideration. In the future, we plan to study the effect of spin currents.
It is also important to note that if the spin degrees of freedom are neglected, this solution goes to zero~\cite{BaMo}.

The present analysis shows that the electrical state considered here has a complex internal structure, 
but invariably demonstrates an overall irrotational flow pattern.

\section{Conclusion}\label{V}

The WFM method was applied to solve TDHFB equations for the harmonic oscillator
with the spin-orbital potential plus
quadrupole-quadrupole and spin-spin residual interactions. 
The dynamics of collective variables
$\L^{\tau\varsigma}_{\lambda\mu}(t)$,
$\R^{\tau\varsigma}_{\lambda\mu}(t)$ and
$\P^{\tau\varsigma}_{\lambda\mu}(t)$ 
with $\lambda=1,2$ and $\mu=0,1,2$ was studied taking into account pair
correlations.

The subject of the especial interest was the predicted low-lying $K^\pi=1^+$ excitation 
located at 1.47~MeV, below of all
scissors modes, and having a $B(E2)$ value of 25.44 W.u., which allowed us to assume the electric character of this state.
It is found that this state is one of three branches of 
$2^+$ state which can exist in spherical nuclei and which is split due to 
the deformation.
An analysis of the dynamical equations allows us to conclude that the variables describing the quadrupole distortions of the shape of the nucleus $\R^{+}_{21}(t)$, 
the Fermi surface deformation $\P^{+}_{21}(t)$, as well as the variables $\R^{\uparrow\downarrow}_{22}(t)$,
$\R^{\downarrow\uparrow}_{20}(t)$, 
$\P^{\uparrow\downarrow}_{22}(t)$,
$\P^{\downarrow\uparrow}_{20}(t)$ associated with spin-flip processes, 
are predominantly responsible for the formation of this excitation.
The underlying physical nature of this mode is revealed by an examination of the nucleons currents.
Despite the complex internal structure of the flow, which is determined by the superposition of irrotational and rotational contributions,
the nucleus as a whole demonstrates the quadrupole character of collective oscillations.
All this confirms the conclusion about the electrical nature of the state under study.

It is discovered, that the antiferromagnetic properties of
nuclei lead to the splitting of the low-lying $2^+$ states already at the zero deformation.
Nuclear antiferromagnetism is expressed in the nonzero equilibrium value of the variable $L^{\tau-}_{10}$ 
and manifests itself as a response of the dynamic mean field to an external perturbation.
It is shown that splitting of $2+$ states in spherical nuclei induced by
the nuclear antiferromagnetism is very similar to the Zeeman splitting in an external uniform magnetic field.


\begin{widetext}

\appendix

\section{Dynamical equations}
\label{AppA}

The set of dynamical equations for isovector variables with $\mu=2$ reads

\begin{eqnarray}\label{IV_mu=2}
\dot { \bar\L}^{+}_{22}&=& 2\frac{i}{\hbar}\mu_N\H\bar\L^{+}_{22} +
\frac{1}{m}\bar\P^+_{22}
-m\,\omega^2 \bar\R^+_{22} -i\hbar\eta\left[ \bar\L^-_{22} + \bar\L^\d_{21} \right]
\nonumber\\
&&+ 2\kappa_0 \left[\left(\left(1+\alpha\right)Q_{20}-2\alpha Q_{00}\right)\bar\R^+_{22}+\left(\left(1+\alpha\right)\bar Q_{20}-2\bar Q_{00}\right)\R^+_{22}\right], 
\nonumber\\
     \dot { \bar\R}^+_{22}&=& 2\frac{i}{\hbar}\mu_N\H\bar\R^{+}_{22} +
\frac{2}{m} \bar\L^+_{22}
-i\hbar\eta\left[\bar\R^-_{22}+\bar\R^\d_{21}\right],
\nonumber
\\   
     \dot { \bar\P}^{+}_{22}&=& 2\frac{i}{\hbar}\mu_N\H\bar\P^{+}_{22} 
-2m\,\omega^2 \bar\L^+_{22}
 + 4\kappa_0\left[Q_{20}\bar\L^+_{22}+\alpha\bar Q_{20}\L^+_{22}\right]
-i\hbar\eta\left[ \bar\P^-_{22} + \bar\P^\d_{21} \right]
\nonumber\\
&&+\frac{3I_2\hbar^2}{2 A_1A_2}\chi \left[\left( \Qx+\Q \right)\bar\L^+_{22}+\left( \bQx+\bQ \right)\L^+_{22}\right]
+\frac{4}{\hbar}\left[I_{pp}^{\kappa\Delta} {\bar{\tilde \P}}_{22}+\bar I_{pp}^{\kappa\Delta} {\tilde \P}_{22}\right]
,\nonumber\\
\dot { \bar\L}^{-}_{22}&=& 2\frac{i}{\hbar}\mu_N\H\bar\L^{-}_{22} +
\frac{1}{m}\bar\P^-_{22}
-m\,\omega^2 \bar\R^-_{22} + 2\kappa_0\left[Q_{20}\bar\R^-_{22}+ \alpha\bar Q_{20}\R^-_{22}\right]
-i\hbar\eta \bar\L^+_{22}
\nonumber\\
&&
+\frac{I_1\hbar^2}{15 A_1A_2} 
\left[\left( 3\chi-\bar\chi \right)\left(\Qx+\Q\right)\bar\R^-_{22}+\left( 3\chi+\bar\chi \right)\left(\bQx+\bQ\right)\R^-_{22}\right]
+\frac{4}{\hbar} \left[I_{rp}^{\kappa\Delta} {\bar{\tilde \L}}_{22}
+\bar I_{rp}^{\kappa\Delta} {{\tilde \L}}_{22}\right]
,\nonumber\\
     \dot { \bar\R}^-_{22}&=& 2\frac{i}{\hbar}\mu_N\H\bar\R^{-}_{22} +
\frac{2}{m} \bar\L^-_{22}
-i\hbar\eta \bar\R^+_{22},
\nonumber
\\   
     \dot { \bar\P}^{-}_{22}&=& 2\frac{i}{\hbar}\mu_N\H\bar\P^{-}_{22} 
-2m\,\omega^2 \bar\L^-_{22}
 + 4\kappa_0\left[Q_{20}\bar\L^-_{22}+ \alpha\bar Q_{20}\L^-_{22}\right]
 - 12\sqrt2 \kappa_0\left[\alpha L^-_{10}(\eq)\bar\R^+_{22}+\bar L^-_{10}(\eq)\R^+_{22}\right]
\nonumber\\
&&-i\hbar\eta \bar\P^+_{22}+\frac{3I_2\hbar^2}{2 A_1A_2}\chi \left[\left( \Qx+\Q \right)\bar\L^-_{22}+\left( \bQx+\bQ \right)\L^-_{22}\right]
,\nonumber
\end{eqnarray}\begin{eqnarray}
     \dot { \bar\L}^{\d}_{21}&=& \frac{i}{\hbar}\mu_N\H(2-g)\bar\L^\d_{21} +
\frac{1}{m} \bar\P^\d_{21}
-\left[m\,\omega^2 + \kappa_0 Q_{20}\right] \bar\R^\d_{21} -\alpha\kappa_0\bar Q_{20} \R^\d_{21}
-i\hbar\frac{\eta}{2} \bar\L^+_{22}\nonumber\\
&&
+\frac{I_1\hbar^2}{15 A_1A_2} 
\left[\left( 3\chi-\bar\chi \right)\left(\Qx+\Q/4\right)\bar\R^\d_{21}+\left( 3\chi+\bar\chi \right)\left(\bQx+\bQ/4\right)\R^\d_{21}\right]
,\nonumber\\
     \dot { \bar\R}^\d_{21}&=& \frac{i}{\hbar}\mu_N\H(2-g)\bar\R^\d_{21} +
\frac{2}{m} \bar\L^\d_{21} - i\hbar\frac{\eta}{2} \bar\R^+_{22},
\nonumber
\\
     \dot { \bar\P}^{\d}_{21}&=& \frac{i}{\hbar}\mu_N\H(2-g)\bar\P^\d_{21}
-2m\,\omega^2 \bar\L^\d_{21} + 2\kappa_0\left[Q_{20}\left(3\bar\L^\d_{11}-\bar\L^\d_{21}\right) + \alpha \bar Q_{20}\left(3\L^\d_{11}-\L^\d_{21}\right)\right]
-i\hbar\frac{\eta}{2} \bar\P^+_{22}
\nonumber\\
&&+\frac{3I_2\hbar^2}{8 A_1A_2}\chi \left[3\Q\bar\L^\d_{11}+\left( 4\Qx+\Q \right)\bar\L^\d_{21}+3\bQ\L^\d_{11}+\left( 4\bQx+\bQ \right)\L^\d_{21}\right]
,\nonumber\\
     \dot { \bar\L}^{\d}_{11}&=& \frac{i}{\hbar}\mu_N\H(2-g)\bar\L^\d_{11}
- 3\kappa_0\left[ Q_{20} \bar\R^\d_{21} + \alpha\bar Q_{20} \R^\d_{21}\right]
-\frac{I_1\hbar^2}{20 A_1A_2} 
\left[\left( 3\chi-\bar\chi \right)\Q\bar\R^\d_{21}+\left( 3\chi+\bar\chi \right)\bQ\R^\d_{21}\right]
,\nonumber\\
     \dot{\bar{\tilde \L}}_{22} &=&  2\frac{i}{\hbar}\mu_N\H{\bar{\tilde\L}}_{22} 
     -\frac{\Delta}{2\hbar} \bar\L^-_{22}-\frac{\bar\Delta}{2\hbar} \L^-_{22},
\nonumber\\
     \dot{\bar{\tilde \P}}_{22} &=& 2\frac{i}{\hbar}\mu_N\H{\bar{\tilde\P}}_{22} 
     -\frac{\Delta}{2\hbar} \bar\P^+_{22}-\frac{\bar\Delta}{2\hbar} \P^+_{22} + 
 3 \hbar\kappa_0 \left[\alpha K_0 \bar\R^+_{22}+\bar K_0 \R^+_{22}\right].
\end{eqnarray}


The set of dynamical equations for isoscalar variables with $\mu=2$ reads

\begin{eqnarray}\label{IS_mu=2}
\dot { \L}^{+}_{22}&=& -2\frac{i}{\hbar}\mu_N\H \L^{+}_{22} +
\frac{1}{m} \P^+_{22}
-m\,\omega^2 \R^+_{22} 
-i\hbar\eta\left[ \L^-_{22} + \L^\d_{21} \right]
\nonumber\\
&&+ 4\kappa_0 \left[\left(Q_{20}-Q_{00}\right)\R^+_{22}+\alpha\left(\bar Q_{20}-\bar Q_{00}\right)\bar\R^+_{22}\right], 
\nonumber
\\
     \dot { \R}^+_{22}&=& -2\frac{i}{\hbar}\mu_N\H \R^{+}_{22} +
\frac{2}{m} \L^+_{22}
-i\hbar\eta\left[\R^-_{22}+\R^\d_{21}\right],
\nonumber
\\   
     \dot { \P}^{+}_{22}&=& -2\frac{i}{\hbar}\mu_N\H \P^{+}_{22} 
-2m\,\omega^2 \L^+_{22}
 + 4\kappa_0\left[Q_{20}\L^+_{22}+\alpha\bar Q_{20}\bar\L^+_{22}\right]
-i\hbar\eta\left[ \P^-_{22} + \P^\d_{21} \right]
\nonumber\\
&&+\frac{3I_2\hbar^2}{2 A_1A_2}\chi \left[\left( \Qx+\Q \right)\bar\L^+_{22}+\left( \bQx+\bQ \right)\L^+_{22}\right]
+\frac{4}{\hbar}\left[I_{pp}^{\kappa\Delta} {{\tilde \P}}_{22}+\bar I_{pp}^{\kappa\Delta} {\bar{\tilde \P}}_{22}\right]
,\nonumber\\
\dot { \L}^{-}_{22}&=& -2\frac{i}{\hbar}\mu_N\H \L^{-}_{22} +
\frac{1}{m}\P^-_{22}
-m\,\omega^2 \R^-_{22} + 2\kappa_0\left[Q_{20}\R^-_{22}+\alpha \bar Q_{20}\bar\R^-_{22}\right]
+\frac{4}{\hbar} \left[I_{rp}^{\kappa\Delta} {{\tilde \L}}_{22}
+\bar I_{rp}^{\kappa\Delta} {\bar{\tilde \L}}_{22}\right]\nonumber\\
&&-i\hbar\eta \L^+_{22}
+\frac{I_1\hbar^2}{15 A_1A_2} 
\left[\left( 3\chi+\bar\chi \right)\left(\Qx+\Q\right)\R^-_{22}+\left( 3\chi-\bar\chi \right)\left(\bQx+\bQ\right)\bar\R^-_{22}\right]
,\nonumber\\
     \dot { \R}^-_{22}&=& -2\frac{i}{\hbar}\mu_N\H \R^{-}_{22} +
\frac{2}{m} \L^-_{22}
-i\hbar\eta \R^+_{22},
\nonumber
\\   
     \dot { \P}^{-}_{22}&=& -2\frac{i}{\hbar}\mu_N\H \P^{-}_{22} 
-2m\,\omega^2 \L^-_{22}
 + 4\kappa_0\left[Q_{20}\L^-_{22}+\alpha \bar Q_{20}\bar\L^-_{22}\right]
 - 12\sqrt2 \kappa_0\left[L^-_{10}(\eq)\R^+_{22}+\alpha\bar L^-_{10}(\eq)\bar\R^+_{22}\right]
\nonumber\\
&&-i\hbar\eta \P^+_{22}+\frac{3I_2\hbar^2}{2 A_1A_2}\chi \left[\left( \Qx+\Q \right)\L^-_{22}+\left( \bQx+\bQ \right)\bar\L^-_{22}\right]
,\nonumber\\
     \dot { \L}^{\d}_{21}&=& -\frac{i}{\hbar}\mu_N\H(2+g) \L^\d_{21} +
\frac{1}{m} \P^\d_{21}
-\left[m\,\omega^2 + \kappa_0 Q_{20}\right] \R^\d_{21} -\alpha\kappa_0\bar Q_{20} \bar\R^\d_{21}
-i\hbar\frac{\eta}{2} \L^+_{22}\nonumber\\
&&
+\frac{I_1\hbar^2}{15 A_1A_2} 
\left[\left( 3\chi+\bar\chi \right)\left(\Qx+\Q/4\right)\R^\d_{21}+\left( 3\chi-\bar\chi \right)\left(\bQx+\bQ/4\right)\bar\R^\d_{21}\right]
,\nonumber\\
     \dot { \R}^\d_{21}&=& -\frac{i}{\hbar}\mu_N\H(2+g) \R^\d_{21} +
\frac{2}{m} \L^\d_{21} - i\hbar\frac{\eta}{2} \R^+_{22},
\nonumber
\\
     \dot { \P}^{\d}_{21}&=& -\frac{i}{\hbar}\mu_N\H(2+g) \P^\d_{21} 
-2m\,\omega^2 \L^\d_{21} + 2\kappa_0\left[Q_{20}\left(3\L^\d_{11}-\L^\d_{21}\right) + \alpha\bar Q_{20}\left(3\bar\L^\d_{11}-\bar\L^\d_{21}\right)\right]
-i\hbar\frac{\eta}{2} \P^+_{22}
\nonumber\\
&&+\frac{3I_2\hbar^2}{8 A_1A_2}\chi \left[3\Q\L^\d_{11}+\left( 4\Qx+\Q \right)\L^\d_{21}+3\bQ\bar\L^\d_{11}+\left( 4\bQx+\bQ \right)\bar\L^\d_{21}\right]
,\nonumber\\
     \dot { \L}^{\d}_{11}&=& -\frac{i}{\hbar}\mu_N\H(2+g) \L^\d_{11} 
- 3\kappa_0\left[ Q_{20} \R^\d_{21} + \alpha\bar Q_{20} \bar\R^\d_{21}\right]
-\frac{I_1\hbar^2}{20 A_1A_2} 
\left[\left( 3\chi+\bar\chi \right)\Q\R^\d_{21}+\left( 3\chi-\bar\chi \right)\bQ\bar\R^\d_{21}\right]
,\nonumber\\
     \dot{{\tilde \L}}_{22} &=& -2\frac{i}{\hbar}\mu_N\H{{\tilde\L}}_{22}
     -\frac{\Delta}{2\hbar} \L^-_{22}-\frac{\bar\Delta}{2\hbar} \bar\L^-_{22},
\nonumber\\
     \dot{{\tilde \P}}_{22} &=& -2\frac{i}{\hbar}\mu_N\H{{\tilde\P}}_{22}
     -\frac{\Delta}{2\hbar} \P^+_{22} -\frac{\bar\Delta}{2\hbar} \bar\P^+_{22} + 
 3 \hbar\kappa_0 \left[K_0 \R^+_{22}+\alpha\bar K_0 \bar\R^+_{22}\right].
\end{eqnarray}

The set of dynamical equations for isovector variables with $\mu=1$ reads

\begin{eqnarray}
     \dot {\bar\L}^{+}_{21}&=& \frac{i}{\hbar}\mu_N\H\bar\L^{+}_{21} +
\frac{1}{m}\bar\P_{21}^{+}
-\left[m\,\omega^2+\kappa_0 \left(4\alpha Q_{00}+(1+\alpha)Q_{20}\right)\right]\bar\R^{+}_{21}
-i\hbar\frac{\eta}{2}\left[\bar\L_{21}^-
+2\bar\L^\u_{22}+
\sqrt6\bar\L^\d_{20}\right]
\nonumber\\&&-
\kappa_0\left(4\bar Q_{00}+(1+\alpha) \bar Q_{20}\right) \R^{+}_{21}
,
\nonumber\\
     \dot {\bar\L}^{-}_{21}&=& \frac{i}{\hbar}\mu_N\H\bar\L^{-}_{21} +
\frac{1}{m}\bar\P_{21}^{-}
-\left[m\,\omega^2+\kappa_0 Q_{20}
-\frac{\hbar^2}{15} \left( 3\chi-\bar\chi \right) \frac{I_1}{A_1A_2} \left(Q_{00}+Q_{20}/4\right)
\right]\bar\R^{-}_{21}-i\hbar\frac{\eta}{2}\bar\L_{21}^+
\nonumber\\
&&
 +\frac{4}{\hbar} I_{rp}^{\kappa\Delta}(r') {\bar{\tilde\L}}_{21} 
-\left[\alpha\kappa_0 \bar Q_{20}
- \frac{\hbar^2}{15}
\left( 3\chi+\bar\chi \right) \frac{I_1}{A_1A_2} \left(\bar Q_{00}+\bar Q_{20}/4\right)\right]\R^{-}_{21}
+\frac{4}{\hbar}\bar I_{rp}^{\kappa\Delta}(r') {{\tilde\L}}_{21}
,
\nonumber\\
     \dot {\bar\L}^\u_{22}&=&
\frac{1}{m}\bar\P_{22}^\u-
\left[m\,\omega^2-2\kappa_0 Q_{20}
-\frac{\hbar^2 }{15}\left( 3\chi-\bar\chi \right)\frac{\sqrt3 I_1}{A_2}
\right]\bar\R^\u_{22}
-i\hbar\frac{\eta}{2}\bar\L_{21}^+
\nonumber\\
&&+
\left[2\alpha\kappa_0 \bar Q_{20} + \frac{\hbar^2 }{15} 
\left( 3\chi+\bar\chi \right)\frac{I_1}{A_1A_2}\left(\bar Q_{20}+\bar Q_{00}\right)
\right]\R^\u_{22}
,
\nonumber\\
\dot {\bar\L}^\d_{20}&=& \frac{i}{\hbar}\mu_N\H(1-g)\bar\L^\d_{20} +
\frac{1}{m}\bar\P_{20}^\d-
\left[m\,\omega^2
+2\kappa_0 Q_{20}\right]\bar\R^\d_{20}
+2\sqrt2\kappa_0 Q_{20}\,\bar\R^\d_{00}
-i\hbar\frac{\eta}{2}\sqrt{\frac{3}{2}}\bar\L_{21}^+ 
\nonumber\\
&&+\frac{\hbar^2}{15} 
\left( 3\chi-\bar\chi \right)\frac{I_1}{A_1A_2} \,
\left[
Q_{00}\bar\R_{20}^\d+
Q_{20}\bar\R_{00}^\d/\sqrt2
\right],
\nonumber\\
&&
-2\alpha\kappa_0 \bar Q_{20}\left[ \R_{20}^\d+\sqrt2\R_{00}^\d\right]
+ \frac{\hbar^2}{15} 
\left( 3\chi+\bar\chi \right)\frac{I_1}{A_1A_2} \,
\left[
\bar Q_{00}\R_{20}^\d+
\bar Q_{20}\R_{00}^\d/\sqrt2
\right]
,
\nonumber\\
\dot {\bar\L}^{+}_{11}&=& \frac{i}{\hbar}\mu_N\H\bar\L^{+}_{11} 
-3(1-\alpha)\kappa_0 Q_{20}\,\bar\R^{+}_{21}
-i\hbar\frac{\eta}{2}\left[\bar\L_{11}^- 
+\sqrt2\bar\L^\d_{10}\right]
+
3(1-\alpha)\kappa_0 \bar Q_{20}\,\R^{+}_{21}
,
\nonumber\\
     \dot {\bar\L}^{-}_{11}&=& \frac{i}{\hbar}\mu_N\H\bar\L^{-}_{11} 
-\left[3\kappa_0 Q_{20}
+\frac{\hbar^2}{20} 
\left( 3\chi-\bar\chi \right)\frac{I_1}{A_1A_2} Q_{20}
\right]\bar\R^{-}_{21}
 +\frac{4}{\hbar} I_{rp}^{\kappa\Delta}(r') {\bar{\tilde\L}}_{11}
 -\hbar\frac{\eta}{2}\left[i\bar\L_{11}^+
+\hbar \bar\F^\d\right]
\nonumber\\
&&
-
\left[3\alpha\kappa_0 \bar Q_{20}
+\frac{\hbar^2}{20}  
\left( 3\chi+\bar\chi \right)\frac{I_1}{A_1A_2}\bar Q_{20}\right]\R^{-}_{21}
+\frac{4}{\hbar}\bar I_{rp}^{\kappa\Delta}(r') {{\tilde\L}}_{11}
,
\nonumber\\
     \dot{\bar\L}^\d_{10}&=& \frac{i}{\hbar}\mu_N\H(1-g)\bar\L^\d_{10} 
-\hbar\frac{\eta}{2\sqrt2}\left[i\bar\L_{11}^+
+\hbar \bar\F^\d\right],
\nonumber\\
     \dot{\bar\F}^\d&=& \frac{i}{\hbar}\mu_N\H(1-g)\bar\F^\d
-\eta\left[\bar\L_{11}^- +\sqrt2\bar\L^\d_{10}\right],
\nonumber\\
\dot {\bar\R}^{+}_{21}&=& \frac{i}{\hbar}\mu_N\H\bar\R^{+}_{21} +
\frac{2}{m}\bar\L_{21}^{+}
-i\hbar\frac{\eta}{2}\left[\bar\R_{21}^-
+2\bar\R^\u_{22}+
\sqrt6\bar\R^\d_{20}\right],
\nonumber\\
     \dot {\bar\R}^{-}_{21}&=& \frac{i}{\hbar}\mu_N\H\bar\R^{-}_{21} +
\frac{2}{m}\bar\L_{21}^{-}
-i\hbar\frac{\eta}{2}\bar\R_{21}^+,
\nonumber\\
     \dot {\bar\R}^\u_{22}&=&
\frac{2}{m}\bar\L_{22}^\u
-i\hbar\frac{\eta}{2}\bar\R_{21}^+,
\nonumber\\
     \dot {\bar\R}^\d_{20}&=& \frac{i}{\hbar}\mu_N\H(1-g)\bar\R^\d_{20} +
\frac{2}{m}\bar\L_{20}^\d
-i\hbar\frac{\eta}{2}\sqrt{\frac{3}{2}}\bar\R_{21}^+,
\nonumber\\
     \dot {\bar\P}^{+}_{21}&=& \frac{i}{\hbar}\mu_N\H\bar\P^{+}_{21} 
-2\left[m\,\omega^2+\kappa_0 Q_{20}\right]\bar\L^{+}_{21}
+6\kappa_0 Q_{20}\bar\L^{+}_{11}
-i\hbar\frac{\eta}{2}\left[\bar\P_{21}^- 
+2\bar\P^\u_{22}+\sqrt6\bar\P^\d_{20}\right]
\nonumber\\
&&+\frac{3}{8} \hbar^2\chi\frac{I_2}{A_1A_2}\left[\left(Q_{20}+4 Q_{00}\right)\bar\L_{21}^{+} +3 Q_{20} \bar\L_{11}^{+}\right]
+\frac{4}{\hbar} I_{pp}^{\kappa\Delta}(r') {\bar{\tilde\P}}_{21}
\nonumber\\
&&+
2\alpha\kappa_0 \bar Q_{20}\left(3\L^{+}_{11}-\L^{+}_{21}\right)
+\frac{3}{8} \hbar^2\chi\frac{I_2}{A_1A_2}\left[\left(\bar Q_{20}+4\bar Q_{00}\right)\L_{21}^{+} +3\bar Q_{20} \L_{11}^{+}\right]
+\frac{4}{\hbar} \bar I_{pp}^{\kappa\Delta}(r') {{\tilde\P}}_{21}
,
\nonumber\\
     \dot {\bar\P}^{-}_{21}&=& \frac{i}{\hbar}\mu_N\H\bar\P^{-}_{21} 
-2\left[m\,\omega^2+\kappa_0 Q_{20}\right]\bar\L^{-}_{21}
+6\kappa_0 Q_{20}\bar\L^{-}_{11}
-6\sqrt2\alpha\kappa_0 L_{10}^-(\eq)\bar\R^{+}_{21}
-i\hbar\frac{\eta}{2}\bar\P_{21}^{+}
\nonumber\\
&&+\frac{3}{8} \hbar^2\chi\frac{I_2}{A_1A_2}\left[\left(Q_{20}+4 Q_{00}\right)\bar\L_{21}^{-} +3 Q_{20} \bar\L_{11}^{-}\right]
+2\alpha\kappa_0 \bar Q_{20}\left(3\L^{-}_{11}-\L^{-}_{21}\right)
\nonumber\\
&&
%
+\frac{3}{8} \hbar^2\chi\frac{I_2}{A_1A_2}\left[\left(\bar Q_{20}+4\bar Q_{00}\right)\L_{21}^{-} +3\bar Q_{20} \L_{11}^{-}\right]
-6\sqrt2\kappa_0 \bar L_{10}^-(\eq)\R^{+}_{21}
,
\nonumber
\end{eqnarray}\begin{eqnarray}
     \dot {\bar\P}^\u_{22}&=&
-2\left[m\,\omega^2-2\kappa_0 Q_{20}\right]\bar\L^\u_{22}
-i\hbar\frac{\eta}{2}\bar\P_{21}^{+}
+\frac{3}{2}\hbar^2 \chi \frac{I_2}{A_1A_2}\left(Q_{20}+Q_{00}\right)
\bar\L^\u_{22}
\nonumber\\
&&+
4\alpha\kappa_0 \bar Q_{20} \L^\u_{22}
+\frac{3}{2}\hbar^2 \chi \frac{I_2}{A_1A_2}\left(\bar Q_{20}+\bar Q_{00}\right)\L^\u_{22}
,
\nonumber\\
     \dot {\bar\P}^\d_{20}&=& \frac{i}{\hbar}\mu_N\H(1-g)\bar\P^\d_{20} 
-2\left[m\,\omega^2+2\kappa_0 Q_{20}\right]\bar\L^\d_{20}
+4\sqrt2\kappa_0 Q_{20}\bar\L^\d_{00}
-i\hbar\frac{\eta}{2}\sqrt{\frac{3}{2}}\bar\P_{21}^{+}
\nonumber\\
&&+\frac{3}{2}\hbar^2 \chi \frac{I_2}{A_1A_2}\left[Q_{00}\bar\L_{20}^\d+Q_{20} \bar\L_{00}^\d/ \sqrt2\right]+
4\alpha\kappa_0\bar Q_{20}\left(\sqrt2\L^\d_{00}-\L^\d_{20}\right)
\nonumber\\&&
+\frac{3}{2}\hbar^2 \chi \frac{I_2}{A_1A_2}\left[\bar Q_{00}\L_{20}^\d+\bar Q_{20} \L_{00}^\d/ \sqrt2\right]
,
\nonumber\\
     \dot {\bar\L}^\d_{00}&=& \frac{i}{\hbar}\mu_N\H(1-g)\bar\L^\d_{00} +
\frac{1}{m}\bar\P_{00}^\d-m\,\omega^2\bar\R^\d_{00}
+2\sqrt2\kappa_0 Q_{20}\bar\R^\d_{20}
\nonumber\\
&&+\frac{\hbar^2}{4\,A_1A_2} 
\left[\left( \chi-\frac{\bar\chi}{3} \right)I_1-\frac{9}{4}\chi I_2\right]
\left[\left(2 Q_{00}+Q_{20}\right)\bar\R_{00}^\d+
\sqrt2 Q_{20}\bar\R_{20}^\d
\right]
\nonumber\\
&&+
2\sqrt2\alpha\kappa_0\bar Q_{20} \R^\d_{20}
+\frac{\hbar^2}{4\,A_1A_2} 
\left[\left( \chi+\frac{\bar\chi}{3} \right)I_1-\frac{9}{4}\chi I_2\right]
\left[\left(2 \bar Q_{00}+\bar Q_{20}\right)\R_{00}^\d+
\sqrt2\bar Q_{20}\R_{20}^\d
\right]
,
\nonumber\\
     \dot {\bar\R}^\d_{00}&=& \frac{i}{\hbar}\mu_N\H(1-g)\bar\R^\d_{00} +
\frac{2}{m}\bar\L_{00}^\d,
\nonumber\\
     \dot {\bar\P}^\d_{00}&=& \frac{i}{\hbar}\mu_N\H(1-g)\bar\P^\d_{00} +
-2m\,\omega^2\bar\L^\d_{00}
+4\sqrt2\kappa_0 Q_{20}\bar\L^\d_{20}
+\frac{3}{4}\hbar^2 
\chi \frac{I_2}{A_1A_2}
\left[\left(2 Q_{00}+Q_{20}\right)\bar \L_{00}^\d+
\sqrt2 Q_{20}\bar \L_{20}^\d
\right]
 \nonumber\\
&&+
4\sqrt2\alpha\kappa_0\bar Q_{20}\L^\d_{20}
+\frac{3}{4}\hbar^2 
\chi \frac{I_2}{A_1A_2}
\left[\left(2 \bar Q_{00}+\bar Q_{20}\right)\L_{00}^\d+
\sqrt2\bar Q_{20}\L_{20}^\d
\right]
,
\nonumber\\ 
     \dot{\bar{\tilde\P}}_{21} &=& \frac{i}{\hbar}\mu_N\H{\bar{\tilde\P}}_{21} 
     -\frac{1}{2\hbar}\Delta(r') \bar\P^+_{21} 
+  6 \hbar\alpha\kappa_0 K_0{\bar\R}^+_{21}
-\frac{1}{2\hbar}\bar \Delta(r') \P^+_{21} + 
 6 \hbar\kappa_0\bar K_0{\R}^+_{21}
, 
\nonumber\\
     \dot{\bar{\tilde\L}}_{21} &=& \frac{i}{\hbar}\mu_N\H{\bar{\tilde\L}}_{21}
     -\frac{1}{2\hbar}\Delta(r') \bar\L^-_{21}-
     \frac{1}{2\hbar}\bar \Delta(r') \L^-_{21}
     ,
\nonumber\\
     \dot{\bar{\tilde\L}}_{11} &=& \frac{i}{\hbar}\mu_N\H{\bar{\tilde\L}}_{11}
     -\frac{1}{2\hbar}\Delta(r') \bar\L^-_{11}-
     \frac{1}{2\hbar}\bar \Delta(r') \L^-_{11}
     .
\label{iv}
\end{eqnarray}

The set of dynamical equations for isoscalar variables with $\mu=1$ reads

\begin{eqnarray}
     \dot {\L}^{+}_{21}&=& -\frac{i}{\hbar}\mu_N\H\L^{+}_{21} +
\frac{1}{m}\P_{21}^{+}
-\left[m\,\omega^2+2\kappa_0 \left(2 Q_{00}+Q_{20}\right)\right]\R^{+}_{21}
-i\hbar\frac{\eta}{2}\left[\L_{21}^-
+2\L^\u_{22}+
\sqrt6\L^\d_{20}\right]
\nonumber\\&&-
\alpha\kappa_0\left(2\bar Q_{00}+ \bar Q_{20}\right)
\bar\R^{+}_{21}
,
\nonumber\\
     \dot {\L}^{-}_{21}&=& -\frac{i}{\hbar}\mu_N\H\L^{-}_{21} +
\frac{1}{m}\P_{21}^{-}
-\left[m\,\omega^2+\kappa_0 Q_{20}
-\frac{\hbar^2}{15} \left( 3\chi+\bar\chi \right) \frac{I_1}{A_1A_2} \left(Q_{00}+Q_{20}/4\right)
\right]\R^{-}_{21}-i\hbar\frac{\eta}{2}\L_{21}^+
\nonumber\\
&&
 +\frac{4}{\hbar} I_{rp}^{\kappa\Delta}(r') {{\tilde\L}}_{21}-
\left[\alpha\kappa_0 \bar Q_{20}
- \frac{\hbar^2}{15}
\left( 3\chi-\bar\chi \right) \frac{I_1}{A_1A_2} \left(\bar Q_{00}+\bar Q_{20}/4\right)\right]\bar\R^{-}_{21}
+\frac{4}{\hbar}\bar I_{rp}^{\kappa\Delta}(r') {\bar{\tilde\L}}_{21}
,
\nonumber\\
     \dot {\L}^\u_{22}&=&
\frac{1}{m}\P_{22}^\u-
\left[m\,\omega^2-2\kappa_0 Q_{20}
-\frac{\hbar^2 }{15} 
\left( 3\chi+\bar\chi \right)\frac{\sqrt3 I_1}{A_2}
\right]\R^\u_{22}
-i\hbar\frac{\eta}{2}\L_{21}^+
\nonumber\\
&&+
\left[2\alpha\kappa_0 \bar Q_{20} + \frac{\hbar^2 }{15} 
\left( 3\chi-\bar\chi \right)\frac{I_1}{A_1A_2}\left(\bar Q_{20}+\bar Q_{00}\right)
\right]\bar\R^\u_{22}
,
\nonumber\\
\dot {\L}^\d_{20}&=& -\frac{i}{\hbar}\mu_N\H(1+g)\L^\d_{20} +
\frac{1}{m}\P_{20}^\d-
\left[m\,\omega^2
+2\kappa_0 Q_{20}\right]\R^\d_{20}
+2\sqrt2\kappa_0 Q_{20}\,\R^\d_{00}
-i\hbar\frac{\eta}{2}\sqrt{\frac{3}{2}}\L_{21}^+ 
\nonumber\\
&&+\frac{\hbar^2}{15} 
\left( 3\chi+\bar\chi \right)\frac{I_1}{A_1A_2} \,
\left[
Q_{00}\R_{20}^\d+
Q_{20}\R_{00}^\d/\sqrt2
\right],
\nonumber\\
&&
-2\alpha\kappa_0 \bar Q_{20}\left[ \bar\R_{20}^\d+\sqrt2\bar\R_{00}^\d\right]
+ \frac{\hbar^2}{15} 
\left( 3\chi-\bar\chi \right)\frac{I_1}{A_1A_2} \,
\left[
\bar Q_{00}\bar\R_{20}^\d+
\bar Q_{20}\bar\R_{00}^\d/\sqrt2
\right]
,
\nonumber
\end{eqnarray}\begin{eqnarray}
\dot {\L}^{+}_{11}&=& -\frac{i}{\hbar}\mu_N\H\L^{+}_{11}
-i\hbar\frac{\eta}{2}\left[\L_{11}^- +\sqrt2\L^\d_{10}\right],
\nonumber\\
     \dot {\L}^{-}_{11}&=& -\frac{i}{\hbar}\mu_N\H\L^{-}_{11}
-\left[3\kappa_0 Q_{20}
+\frac{\hbar^2}{20} 
\left( 3\chi+\bar\chi \right)\frac{I_1}{A_1A_2} Q_{20}
\right]\R^{-}_{21}
 +\frac{4}{\hbar} I_{rp}^{\kappa\Delta}(r') {{\tilde\L}}_{11}
 -\hbar\frac{\eta}{2}\left[i\L_{11}^+
+\hbar \F^\d\right]
\nonumber\\
&&
-
\left[3\alpha\kappa_0 \bar Q_{20}
+\frac{\hbar^2}{20}  
\left( 3\chi-\bar\chi \right)\frac{I_1}{A_1A_2}\bar Q_{20}\right]\bar\R^{-}_{21}
+\frac{4}{\hbar}\bar I_{rp}^{\kappa\Delta}(r') {\bar{\tilde\L}}_{11}
,
\nonumber\\
     \dot{\L}^\d_{10}&=& -\frac{i}{\hbar}\mu_N\H(1+g)\L^\d_{10}
-\hbar\frac{\eta}{2\sqrt2}\left[i\L_{11}^+
+\hbar \F^\d\right],
\nonumber\\
     \dot{\F}^\d&=& -\frac{i}{\hbar}\mu_N\H(1+g)\F^\d
-\eta\left[\L_{11}^- +\sqrt2\L^\d_{10}\right],
\nonumber\\
\dot {\R}^{+}_{21}&=& -\frac{i}{\hbar}\mu_N\H\R^{+}_{21}
\frac{2}{m}\L_{21}^{+}
-i\hbar\frac{\eta}{2}\left[\R_{21}^-
+2\R^\u_{22}+
\sqrt6\R^\d_{20}\right],
\nonumber\\
     \dot {\R}^{-}_{21}&=& -\frac{i}{\hbar}\mu_N\H\R^{-}_{21}
\frac{2}{m}\L_{21}^{-}
-i\hbar\frac{\eta}{2}\R_{21}^+,
\nonumber\\
     \dot {\R}^\u_{22}&=&
\frac{2}{m}\L_{22}^\u
-i\hbar\frac{\eta}{2}\R_{21}^+,
\nonumber\\
     \dot {\R}^\d_{20}&=& -\frac{i}{\hbar}\mu_N\H(1+g)\R^\d_{20} +
\frac{2}{m}\L_{20}^\d
-i\hbar\frac{\eta}{2}\sqrt{\frac{3}{2}}\R_{21}^+,
\nonumber\\
     \dot {\P}^{+}_{21}&=& -\frac{i}{\hbar}\mu_N\H\P^{+}_{21}
-2\left[m\,\omega^2+\kappa_0 Q_{20}\right]\L^{+}_{21}
+6\kappa_0 Q_{20}\L^{+}_{11}
-i\hbar\frac{\eta}{2}\left[\P_{21}^- 
+2\P^\u_{22}+\sqrt6\P^\d_{20}\right]
\nonumber\\
&&+\frac{3}{8} \hbar^2\chi\frac{I_2}{A_1A_2}\left[\left(Q_{20}+4 Q_{00}\right)\L_{21}^{+} +3 Q_{20} \L_{11}^{+}\right]
+\frac{4}{\hbar} I_{pp}^{\kappa\Delta}(r') {{\tilde\P}}_{21}
\nonumber\\
&&+
2\alpha\kappa_0 \bar Q_{20}\left(3\bar\L^{+}_{11}-\bar\L^{+}_{21}\right)
+\frac{3}{8} \hbar^2\chi\frac{I_2}{A_1A_2}\left[\left(\bar Q_{20}+4\bar Q_{00}\right)\bar\L_{21}^{+} +3\bar Q_{20} \bar\L_{11}^{+}\right]
+\frac{4}{\hbar} \bar I_{pp}^{\kappa\Delta}(r') {\bar{\tilde\P}}_{21}
,
\nonumber\\
     \dot {\P}^{-}_{21}&=& -\frac{i}{\hbar}\mu_N\H\P^-_{21}
-2\left[m\,\omega^2+\kappa_0 Q_{20}\right]\L^{-}_{21}
+6\kappa_0 Q_{20}\L^{-}_{11}
-6\sqrt2\kappa_0 L_{10}^-(\eq)\R^{+}_{21}
-i\hbar\frac{\eta}{2}\P_{21}^{+}
\nonumber\\
&&+\frac{3}{8} \hbar^2\chi\frac{I_2}{A_1A_2}\left[\left(Q_{20}+4 Q_{00}\right)\L_{21}^{-} +3 Q_{20} \L_{11}^{-}\right]
\nonumber\\
&&+
2\alpha\kappa_0 \bar Q_{20}\left(3\bar\L^{-}_{11}-\bar\L^{-}_{21}\right)
+\frac{3}{8} \hbar^2\chi\frac{I_2}{A_1A_2}\left[\left(\bar Q_{20}+4\bar Q_{00}\right)\bar\L_{21}^{-} +3\bar Q_{20} \bar\L_{11}^{-}\right]
-6\sqrt2\alpha\kappa_0 \bar L_{10}^-(\eq)\bar\R^{+}_{21}
,
\nonumber\\
     \dot {\P}^\u_{22}&=&
-2\left[m\,\omega^2-2\kappa_0 Q_{20}\right]\L^\u_{22}
-i\hbar\frac{\eta}{2}\P_{21}^{+}
+\frac{3}{2}\hbar^2 \chi \frac{I_2}{A_1A_2}\left(Q_{20}+Q_{00}\right)
\L^\u_{22}
\nonumber\\
&&+
4\alpha\kappa_0 \bar Q_{20} \bar\L^\u_{22}
+\frac{3}{2}\hbar^2 \chi \frac{I_2}{A_1A_2}\left(\bar Q_{20}+\bar Q_{00}\right)\bar\L^\u_{22}
,
\nonumber\\
     \dot{\P}^\d_{20}&=& -\frac{i}{\hbar}\mu_N\H(1+g)\P^\d_{20}
-2\left[m\,\omega^2+2\kappa_0 Q_{20}\right]\L^\d_{20}
+4\sqrt2\kappa_0 Q_{20}\L^\d_{00}
-i\hbar\frac{\eta}{2}\sqrt{\frac{3}{2}}\P_{21}^{+}
\nonumber\\&&
+\frac{3}{2}\hbar^2 \chi \frac{I_2}{A_1A_2}\left[Q_{00}\L_{20}^\d+Q_{20} \L_{00}^\d/ \sqrt2\right]
+4\alpha\kappa_0\bar Q_{20}\left(\sqrt2\bar\L^\d_{00}-\bar\L^\d_{20}\right)\nonumber\\
&&
+\frac{3}{2}\hbar^2 \chi \frac{I_2}{A_1A_2}\left[\bar Q_{00}\L_{20}^\d+\bar Q_{20} \bar\L_{00}^\d/ \sqrt2\right]
,
\nonumber\\
     \dot {\L}^\d_{00}&=& -\frac{i}{\hbar}\mu_N\H(1+g)\L^\d_{00} +
\frac{1}{m}\P_{00}^\d-m\,\omega^2\R^\d_{00}
+2\sqrt2\kappa_0 Q_{20}\R^\d_{20}\nonumber\\
&&
+\frac{\hbar^2}{4\,A_1A_2} 
\left[\left( \chi+\frac{\bar\chi}{3} \right)I_1-\frac{9}{4}\chi I_2\right]
\left[\left(2 Q_{00}+Q_{20}\right)\R_{00}^\d+
\sqrt2 Q_{20}\R_{20}^\d
\right]
\nonumber\\
&&+
2\sqrt2\alpha\kappa_0\bar Q_{20} \bar\R^\d_{20}
+\frac{\hbar^2}{4\,A_1A_2} 
\left[\left( \chi-\frac{\bar\chi}{3} \right)I_1-\frac{9}{4}\chi I_2\right]
\left[\left(2 \bar Q_{00}+\bar Q_{20}\right)\bar\R_{00}^\d+
\sqrt2\bar Q_{20}\bar\R_{20}^\d
\right]
,
\nonumber\\
     \dot {\R}^\d_{00}&=& -\frac{i}{\hbar}\mu_N\H(1+g)\R^\d_{00} +
\frac{2}{m}\L_{00}^\d,
\nonumber\\
     \dot {\P}^\d_{00}&=& -\frac{i}{\hbar}\mu_N\H(1+g)\P^\d_{00} 
-2m\,\omega^2\L^\d_{00}
+4\sqrt2\kappa_0 Q_{20}\L^\d_{20}
+\frac{3}{4}\hbar^2 
\chi \frac{I_2}{A_1A_2}
\left[\left(2 Q_{00}+Q_{20}\right)\bar \L_{00}^\d+
\sqrt2 Q_{20}\bar \L_{20}^\d
\right]
 \nonumber\\
&&+
4\sqrt2\alpha\kappa_0\bar Q_{20}\bar\L^\d_{20}
+\frac{3}{4}\hbar^2 
\chi \frac{I_2}{A_1A_2}
\left[\left(2 \bar Q_{00}+\bar Q_{20}\right)\bar\L_{00}^\d+
\sqrt2\bar Q_{20}\bar\L_{20}^\d
\right]
,
\nonumber\\ 
     \dot{{\tilde\P}}_{21} &=&  -\frac{i}{\hbar}\mu_N\H \tilde\P_{21}
     -\frac{1}{2\hbar}\Delta(r') \P^+_{21} + 
 6 \hbar\kappa_0 K_0\R^+_{21}
-\frac{1}{2\hbar}\bar \Delta(r') \bar\P^+_{21} + 
 6 \hbar\alpha\kappa_0\bar K_0\bar\R^+_{21}
, 
\nonumber\\
     \dot{{\tilde\L}}_{21} &=&  -\frac{i}{\hbar}\mu_N\H \tilde\L_{21}
     -\frac{1}{2\hbar}\Delta(r') \L^-_{21}-
     \frac{1}{2\hbar}\bar \Delta(r') \bar\L^-_{21}
     ,
\nonumber\\
     \dot{{\tilde\L}}_{11} &=&  -\frac{i}{\hbar}\mu_N\H \tilde\L_{11}
     -\frac{1}{2\hbar}\Delta(r') \L^-_{11}-
     \frac{1}{2\hbar}\bar \Delta(r') \bar\L^-_{11}
     .
\label{is}\end{eqnarray}

The set of dynamical equations for isovector variables with $\mu=0$ reads

\begin{eqnarray}\label{IV_mu=0}
\dot {\bar\L}^{+}_{20}&=&
\frac{1}{m}\P^+_{20}
-\left[m\,\omega^2+2\kappa_0\left((1+\alpha)\Q+2\alpha\Qx\right)\right]\bar\R^+_{20} 
+ 2\sqrt2\kappa_0\Q\bar\R^+_{00}  -i\hbar\frac{\eta}{2}\sqrt{6}\left[\bar\L^\u_{21} +\bar\L^\d_{2-1} \right]
\nonumber\\
&&- 2\kappa_0 \left[\left((1+\alpha)\bQ+2\bQx\right)\R^+_{20}-\sqrt2\alpha\bQ\R^+_{00}\right], 
\nonumber\\
\dot {\bar\R}^+_{20}&=&
\frac{2}{m}\bar\L^+_{20}
-i\hbar\frac{\eta}{2}\sqrt{6} \left[\R^\u_{21}+\R^\d_{2-1}\right],
\nonumber\\   
     \dot {\bar\P}^{+}_{20}&=&
-2\left[m\,\omega^2 + 2\kappa_0\Q\right]\bar\L^+_{20}
+4\sqrt2\kappa_0\Q\bar\L^+_{00}
 - 4\alpha\kappa_0 \bQ\left[\L^+_{20}-\sqrt2\L^+_{00}\right]
 \nonumber\\
&&-i\hbar\frac{\eta}{2}\sqrt{6}\left[\bar\P^\u_{21} +\bar\P^\d_{2-1} \right]
+\frac{4}{\hbar}\left[I_{pp}^{\kappa\Delta} {\bar{\tilde \P}}_{20}+\bar I_{pp}^{\kappa\Delta} {{\tilde \P}}_{20}
\right] 
\nonumber\\&&
+\frac{3I_2\hbar^2}{2A_1A_2}\chi\left[\Q\bar\L^+_{00}/\sqrt2+\Qx\bar\L^+_{20}+\bQ\L^+_{00}/\sqrt2+\bQx\L^+_{20}\right]
,\nonumber\\
\dot {\bar\L}^{+}_{10}&=&
-i\hbar\frac{\eta}{2}\sqrt2\left[\bar\L^\u_{11} +\bar\L^\d_{1-1}\right]
,\nonumber\\
     \dot {\bar\L}^{\u}_{11}&=& \frac{i}{\hbar}\mu_N\H\bar\L^\u_{11} +
\hbar\frac{\eta}{4}\left[\hbar\bar\F^- - i\sqrt2\bar\L^+_{10}\right]
-3\,\kappa_0\!\left[\Q \bar\R^\u_{21}+\alpha\bQ \R^\u_{21}\right]
\nonumber\\&&
-\frac{I_1\hbar^2}{20 A_1A_2} 
\left[\left( 3\chi-\bar\chi \right)Q_{20}\bar\R^\u_{21}+\left( 3\chi+\bar\chi \right)\bQ\R^\u_{21}\right]
,\nonumber\\
     \dot {\bar\L}^{\d}_{1-1}&=& -\frac{i}{\hbar}\mu_N\H\bar\L^\d_{1-1} +
\hbar\frac{\eta}{4}\left[\hbar\bar\F^- - i\sqrt2\bar\L^+_{10}\right]
+3\,\kappa_0\!\left[\Q \bar\R^\d_{2-1}+\alpha\bQ \R^\d_{2-1}\right]
\nonumber\\&&
+\frac{I_1\hbar^2}{20 A_1A_2} 
\left[\left( 3\chi-\bar\chi \right)Q_{20}\bar\R^\d_{2-1}+\left( 3\chi+\bar\chi \right)\bQ\R^\d_{2-1}\right]
,\nonumber\\
     \dot{\bar\F}^-&=&
2\eta\left[\bar\L^\u_{11} +\bar\L^\d_{1-1}\right],
\nonumber\\
     \dot {\bar\L}^{\u}_{21}&=& \frac{i}{\hbar}\mu_N\H\bar\L^\u_{21} +
\frac{1}{m}\bar\P^\u_{21}
-\left[m\,\omega^2 + \kappa_0\Q \right]\bar\R^\u_{21} - \alpha\kappa_0\bQ\R^\u_{21}
-i\hbar\frac{\eta}{4}\sqrt{6}\bar\L^+_{20}
\nonumber\\&&
+\frac{I_1\hbar^2}{15 A_1A_2} 
\left[\left( 3\chi-\bar\chi \right)\left(\Qx+\Q/4\right)\bar\R^\u_{21}+\left( 3\chi+\bar\chi \right)\left(\bQx+\bQ/4\right)\R^\u_{21}\right]
,\nonumber\\
     \dot {\bar\L}^{\d}_{2-1}&=& -\frac{i}{\hbar}\mu_N\H\bar\L^\d_{2-1} +
\frac{1}{m}\bar\P^\d_{2-1}
-\left[m\,\omega^2 + \kappa_0\Q \right]\bar\R^\d_{2-1} - \alpha\kappa_0\bQ \R^\d_{2-1}
-i\hbar\frac{\eta}{4}\sqrt{6}\bar\L^+_{20}
\nonumber\\&&
+\frac{I_1\hbar^2}{15 A_1A_2} 
\left[\left( 3\chi-\bar\chi \right)\left(\Qx+\Q/4\right)\bar\R^\d_{2-1}+\left( 3\chi+\bar\chi \right)\left(\bQx+\bQ/4\right)\R^\d_{2-1}\right]
,\nonumber\\
     \dot {\bar\R}^\u_{21}&=& \frac{i}{\hbar}\mu_N\H\bar\R^\u_{21} +
\frac{2}{m}\bar\L^\u_{21} - i\hbar\frac{\eta}{4}\sqrt{6}\bar\R^+_{20},
\nonumber\\
\dot {\bar\R}^\d_{2-1}&=& -\frac{i}{\hbar}\mu_N\H\bar\R^\d_{2-1} +
\frac{2}{m}\bar\L^\d_{2-1} - i\hbar\frac{\eta}{4}\sqrt{6}\bar\R^+_{20},
\nonumber\\
\dot {\bar\P}^{\u}_{21}&=& \frac{i}{\hbar}\mu_N\H\bar\P^\u_{21} 
-2\left[m\,\omega^2 +\kappa_0\Q\right]\bar\L^\u_{21}+6\kappa_0\Q\bar\L^\u_{11}
-i\hbar\frac{\eta}{4}\sqrt{6}\bar\P^+_{20}
 -2\kappa_0 \alpha\bQ\left(\L^\u_{21}-3\L^\u_{11}\right)
\nonumber\\&&
+\frac{3I_2\hbar^2}{8A_1A_2}\chi\left[3\Q\bar\L^\u_{11}+\left(4\Qx+\Q\right)\bar\L^\u_{21}+3\bQ\L^\u_{11}+\left(4\bQx+\bQ\right)\L^\u_{21}\right]
,\nonumber\\
\dot {\bar\P}^{\d}_{2-1}&=& -\frac{i}{\hbar}\mu_N\H\bar\P^\d_{2-1} 
-2\left[m\,\omega^2 +\kappa_0\Q\right]\bar\L^\d_{2-1}-6\kappa_0\Q\bar\L^\d_{1-1}
-i\hbar\frac{\eta}{4}\sqrt{6}\bar\P^+_{20}
-2\kappa_0\alpha\bQ\left(\L^\d_{2-1}+3\L^\d_{1-1}\right)
 \nonumber\\&&
-\frac{3I_2\hbar^2}{8A_1A_2}\chi\left[3\Q\bar\L^\d_{1-1}-\left(4\Qx+\Q\right)\bar\L^\d_{2-1}+3\bQ\L^\d_{1-1}-\left(4\bQx+\bQ\right)\L^\d_{2-1}\right]
,\nonumber\\
\dot{\bar{\tilde \P}}_{20} &=& -\frac{\Delta}{2\hbar} \bar \P^+_{20}-\frac{\bar\Delta}{2\hbar}\P^+_{20}
+ 
 6\kappa_0 \hbar \left[\alpha K_0 \bar \R^+_{20} + \bar K_0 \R^+_{20}
 \right],
\nonumber\\
     \dot {\bar\L}^{+}_{00}&=&
\frac{1}{m}\P^+_{00}
-m\,\omega^2\bar\R^+_{00} + 2\sqrt2(1+\alpha)\kappa_0\left[\Q\bar\R^+_{20}+\bQ\R^+_{20}\right]
\nonumber\\&&\qquad\quad
-\frac{9I_2\hbar^2}{16 A_1A_2}\chi 
\left[\left(2\Qx+\Q\right)\bar\R^+_{00}+\sqrt2\Q\bar\R^+_{20}+\left(2\bQx+\bQ\right)\R^+_{00}+\sqrt2\bQ\R^+_{20}\right]
,\nonumber\\
     \dot {\bar\R}^+_{00}&=&
\frac{2}{m}\bar\L^+_{00},
\nonumber
\end{eqnarray}\begin{eqnarray}  
     \dot {\bar\P}^{+}_{00}&=&
-2m\,\omega^2\bar\L^+_{00}
 + 4\sqrt2\kappa_0\left[\Q\bar\L^+_{20} + \alpha\bQ\L^+_{20}\right]
 \nonumber\\&&
+\frac{3I_2\hbar^2}{2A_1A_2}\chi\left[\Q\bar\L^+_{20}/\sqrt2+\left(\Qx+\Q/2\right)\bar\L^+_{00} + 
\bQ\L^+_{20}/\sqrt2+\left(\bQx+\bQ/2\right)\L^+_{00}\right],
\nonumber\\
     \dot{\bar\L}^{-}_{20}&=&
\frac{1}{m} \bar\P^-_{20}
- \left[m\,\omega^2+2\kappa_0\Q\right] \bar\R^-_{20} +2\sqrt2\kappa_0\Q \bar\R^-_{00}
- 2\alpha\kappa_0\bQ \left[\R^-_{20}-\sqrt2 \R^-_{00}\right]
+\frac{4}{\hbar} \left[I_{rp}^{\kappa\Delta} {\bar{\tilde \L}}_{20}
+\bar I_{rp}^{\kappa\Delta} {{\tilde \L}}_{20}\right]
\nonumber\\&&\qquad\quad
+\frac{I_1\hbar^2}{15 A_1A_2}
\left[\left(3\chi-\bar\chi\right)
\left(\Qx\bar\R^-_{20}+\Q\bar\R^-_{00}/\sqrt2\right)
+\left(3\chi+\bar\chi\right)
\left(\bQx\R^-_{20}+\bQ\R^-_{00}/\sqrt2\right)\right]
,\nonumber\\
     \dot{\bar\R}^-_{20}&=&
\frac{2}{m} \bar\L^-_{20},
\nonumber\\
\dot {\bar\P}^{-}_{20}&=&
-2\left[m\,\omega^2 + 2\kappa_0\Q\right] \bar\L^-_{20} +4\sqrt2\kappa_0\Q\bar\L^-_{00} 
- 4\alpha\kappa_0\bQ\left[\L^-_{20}-\sqrt2 \L^-_{00}\right]
\nonumber\\&&
+\frac{3I_2\hbar^2}{2A_1A_2}\chi\left[\Q\bar\L^-_{00}/\sqrt2+\Qx\bar\L^-_{20}+\bQ\L^-_{00}/\sqrt2+\bQx\L^-_{20}\right]
,\nonumber\\
     \dot{\bar\L}^{-}_{00}&=&
\frac{1}{m} \bar\P^-_{00}
- m\,\omega^2 \bar\R^-_{00} + 2\sqrt2\kappa_0\left[\Q \bar\R^-_{20} + \alpha\bQ \R^-_{20}\right]
\nonumber\\&&\qquad\quad
+\frac{\hbar^2}{12 A_1A_2}
\left[\left(\left(3\chi-\bar\chi\right)I_1-27\chi I_2/4\right)
\left(\left(2\Qx+\Q\right)\bar\R^-_{00}+\sqrt2\Q\bar\R^-_{20}\right)\right.
\nonumber\\&&\qquad\qquad\qquad\quad
\left.+\left(\left(3\chi+\bar\chi\right)I_1-27\chi I_2/4\right)
\left(\left(2\bQx+\bQ\right)\R^-_{00}+\sqrt2\bQ\R^-_{20}\right)\right]
,\nonumber\\
     \dot{\bar\R}^-_{00}&=&
\frac{2}{m} \bar\L^-_{00},
\nonumber\\
\dot{\bar\P}^{-}_{00}&=&
-2m\,\omega^2 \bar\L^-_{00} + 4\sqrt2\kappa_0\left[\Q \bar\L^-_{20}+\alpha\bQ \L^-_{20}\right] 
\nonumber\\&&
+\frac{3I_2\hbar^2}{2A_1A_2}\chi\left[\Q\bar\L^-_{20}/\sqrt2+\left(\Qx+\Q/2\right)\bar\L^-_{00} + 
\bQ\L^-_{20}/\sqrt2+\left(\bQx+\bQ/2\right)\L^-_{00}\right]
,\nonumber\\
 \dot{\bar{\tilde \L}}_{20} &=& -\frac{\Delta}{2\hbar} \bar\L^-_{20}-\frac{\bar\Delta}{2\hbar} \L^-_{20},
\nonumber\\
     \dot{\bar\L}^{-}_{10}&=&\frac{4}{\hbar} \left[I_{rp}^{\kappa\Delta} {\bar{\tilde \L}}_{10}
+\bar I_{rp}^{\kappa\Delta} {{\tilde \L}}_{10}\right],
\nonumber\\
 \dot{\bar{\tilde \L}}_{10} &=& -\frac{\Delta}{2\hbar} \bar \L^-_{10}-\frac{\bar\Delta}{2\hbar} \L^-_{10}.
\end{eqnarray}

The set of dynamical equations for isoscalar variables with $\mu=0$ reads

\begin{eqnarray}\label{IS_mu=0}
\dot {\L}^{+}_{20}&=&
\frac{1}{m}\P^+_{20}
-\left[m\,\omega^2+4\kappa_0\left(\Q+\Qx\right)\right]\R^+_{20} 
+ 2\sqrt2\kappa_0\Q\R^+_{00}  -i\hbar\frac{\eta}{2}\sqrt{6}\left[\L^\u_{21} +\L^\d_{2-1} \right]
\nonumber\\
&&- 2\sqrt2\alpha\kappa_0 \left[\sqrt2\left(\bQ+\bQx\right)\bar\R^+_{20}-\bQ\bar\R^+_{00}\right], 
\nonumber\\
     \dot {\R}^+_{20}&=&
\frac{2}{m}\L^+_{20}
-i\hbar\frac{\eta}{2}\sqrt{6} \left[\R^\u_{21}+\R^\d_{2-1}\right],
\nonumber\\   
     \dot {\P}^{+}_{20}&=&
-2\left[m\,\omega^2 + 2\kappa_0\Q\right]\L^+_{20}
+4\sqrt2\kappa_0\Q\L^+_{00}
 - 4\alpha\kappa_0 \bQ\left[\bar\L^+_{20}-\sqrt2\bar\L^+_{00}\right]
\nonumber\\
&&-i\hbar\frac{\eta}{2}\sqrt{6}\left[\P^\u_{21} +\P^\d_{2-1} \right]
+\frac{4}{\hbar}\left[I_{pp}^{\kappa\Delta} {{\tilde \P}}_{20}+\bar I_{pp}^{\kappa\Delta} {\bar{\tilde \P}}_{20}
\right]
\nonumber\\&&
+\frac{3I_2\hbar^2}{2A_1A_2}\chi\left[\Q\L^+_{00}/\sqrt2+\Qx\L^+_{20}+\bQ\bar\L^+_{00}/\sqrt2+\bQx\bar\L^+_{20}\right]
,\nonumber\\
\dot {\L}^{+}_{10}&=&
-i\hbar\frac{\eta}{2}\sqrt2\left[\L^\u_{11} +\L^\d_{1-1}\right]
,\nonumber\\
     \dot {\L}^{\u}_{11}&=& \frac{i}{\hbar}\mu_N\H\L^\u_{11} +
\hbar\frac{\eta}{4}\left[\hbar\F^- - i\sqrt2\L^+_{10}\right]
-3\kappa_0\!\left[\Q \R^\u_{21}+\alpha\bQ \bar\R^\u_{21}\right]
\nonumber\\&&
-\frac{I_1\hbar^2}{20 A_1A_2} 
\left[\left( 3\chi+\bar\chi \right)Q_{20}\R^\u_{21}+\left( 3\chi-\bar\chi \right)\bQ\bar\R^\u_{21}\right]
,\nonumber\\
     \dot {\L}^{\d}_{1-1}&=& -\frac{i}{\hbar}\mu_N\H\L^\d_{1-1} +
\hbar\frac{\eta}{4}\left[\hbar\F^- - i\sqrt2\L^+_{10}\right]
+3\kappa_0\!\left[\Q \R^\d_{2-1}+\alpha\bQ \bar\R^\d_{2-1}\right]
\nonumber\\&&
+\frac{I_1\hbar^2}{20 A_1A_2} 
\left[\left( 3\chi+\bar\chi \right)Q_{20}\R^\d_{2-1}+\left( 3\chi-\bar\chi \right)\bQ\bar\R^\d_{2-1}\right],
\nonumber
\end{eqnarray}\begin{eqnarray}
     \dot{\F}^-&=&
2\eta\left[\L^\u_{11} +\L^\d_{1-1}\right],
\nonumber\\
     \dot {\L}^{\u}_{21}&=& \frac{i}{\hbar}\mu_N\H\L^\u_{21} +
\frac{1}{m}\P^\u_{21}
-\left[m\,\omega^2 + \kappa_0\Q \right]\R^\u_{21} - \alpha\kappa_0\bQ \bar\R^\u_{21}
-i\hbar\frac{\eta}{4}\sqrt{6}\L^+_{20}
\nonumber\\&&
+\frac{I_1\hbar^2}{15 A_1A_2} 
\left[\left( 3\chi+\bar\chi \right)\left(\Qx+\Q/4\right)\R^\u_{21}+\left( 3\chi-\bar\chi \right)\left(\bQx+\bQ/4\right)\bar\R^\u_{21}\right]
,\nonumber\\
     \dot {\L}^{\d}_{2-1}&=& -\frac{i}{\hbar}\mu_N\H\L^\d_{2-1} +
\frac{1}{m}\P^\d_{2-1}
-\left[m\,\omega^2 + \kappa_0\Q \right]\R^\d_{2-1} - \alpha\kappa_0\bQ \bar\R^\d_{2-1}
-i\hbar\frac{\eta}{4}\sqrt{6}\L^+_{20}
\nonumber\\&&
+\frac{I_1\hbar^2}{15 A_1A_2} 
\left[\left( 3\chi+\bar\chi \right)\left(\Qx+\Q/4\right)\R^\d_{2-1}+\left( 3\chi-\bar\chi \right)\left(\bQx-\bQ/4\right)\bar\R^\d_{2-1}\right]
,\nonumber\\
     \dot {\R}^\u_{21}&=& \frac{i}{\hbar}\mu_N\H\R^\u_{21} +
\frac{2}{m}\L^\u_{21} - i\hbar\frac{\eta}{4}\sqrt{6}\R^+_{20},
\nonumber\\
\dot {\R}^\d_{2-1}&=& -\frac{i}{\hbar}\mu_N\H\R^\d_{2-1} +
\frac{2}{m}\L^\d_{2-1} - i\hbar\frac{\eta}{4}\sqrt{6}\R^+_{20},
\nonumber\\
\dot {\P}^{\u}_{21}&=& \frac{i}{\hbar}\mu_N\H\P^\u_{21} 
-2\left[m\,\omega^2 +\kappa_0\Q\right]\L^\u_{21}+6\kappa_0\Q\L^\u_{11}
-i\hbar\frac{\eta}{4}\sqrt{6}\P^+_{20}
-2\alpha\kappa_0
\bQ\left(\bar\L^\u_{21}-3\bar\L^\u_{11}\right)
 \nonumber\\&&
+\frac{3I_2\hbar^2}{8A_1A_2}\chi\left[3\Q\L^\u_{11}+\left(4\Qx+\Q\right)\L^\u_{21}+3\bQ\bar\L^\u_{11}+\left(4\bQx+\bQ\right)\bar\L^\u_{21}\right]
,\nonumber\\
\dot {\P}^{\d}_{2-1}&=& -\frac{i}{\hbar}\mu_N\H\P^\d_{2-1} 
-2\left[m\,\omega^2 +\kappa_0\Q\right]\L^\d_{2-1}-6\kappa_0\Q\L^\d_{1-1}
-i\hbar\frac{\eta}{4}\sqrt{6}\P^+_{20}
 -2\alpha\kappa_0\bQ\left(\bar\L^\d_{2-1}+3\bar\L^\d_{1-1}\right)
 \nonumber\\&&
-\frac{3I_2\hbar^2}{8A_1A_2}\chi\left[3\Q\L^\d_{1-1}-\left(4\Qx+\Q\right)\L^\d_{2-1}+3\bQ\bar\L^\d_{1-1}-\left(4\bQx+\bQ\right)\bar\L^\d_{2-1}\right]
,\nonumber\\
\dot{{\tilde \P}}_{20} &=& -\frac{\Delta}{2\hbar} \P^+_{20}-\frac{\bar\Delta}{2\hbar}\bar \P^+_{20}
+ 
 6\kappa_0 \hbar \left[K_0 \R^+_{20} + \alpha \bar K_0 \bar \R^+_{20}
 \right],\nonumber\\
      \dot {\L}^{+}_{00}&=&
\frac{1}{m}\P^+_{00}
-m\,\omega^2\R^+_{00} + 4\sqrt2\kappa_0\left[\Q\R^+_{20}+\alpha\bQ\bar\R^+_{20}\right]
\nonumber\\&&\qquad\quad
-\frac{9I_2\hbar^2}{16 A_1A_2}\chi 
\left[\left(2\Qx+\Q\right)\R^+_{00}+\sqrt2\Q\R^+_{20}+\left(2\bQx+\bQ\right)\bar\R^+_{00}+\sqrt2\bQ\bar\R^+_{20}\right]
,\nonumber\\
     \dot {\R}^+_{00}&=&
\frac{2}{m}\L^+_{00},
\nonumber\\   
     \dot {\P}^{+}_{00}&=&
-2m\,\omega^2\L^+_{00}
 + 4\sqrt2\kappa_0\left[\Q\L^+_{20} + \alpha\bQ\bar\L^+_{20}\right] 
\nonumber\\&&
+\frac{3I_2\hbar^2}{2A_1A_2}\chi\left[\Q\L^+_{20}/\sqrt2+\left(\Qx+\Q/2\right)\L^+_{00} + 
\bQ\bar\L^+_{20}/\sqrt2+\left(\bQx+\bQ/2\right)\bar\L^+_{00}\right],
\nonumber\\
     \dot{\L}^{-}_{20}&=&
\frac{1}{m} \P^-_{20}
- \left[m\,\omega^2+2\kappa_0\Q\right] \R^-_{20} +2\sqrt2\kappa_0\Q \R^-_{00} 
- 2\alpha\kappa_0\bQ \left[\bar\R^-_{20}-\sqrt2 \bar\R^-_{00}\right]
+\frac{4}{\hbar} \left[I_{rp}^{\kappa\Delta} {{\tilde \L}}_{20}
+\bar I_{rp}^{\kappa\Delta} {\bar{\tilde \L}}_{20}\right]
\nonumber\\&&\qquad\quad
+\frac{I_1\hbar^2}{15 A_1A_2}
\left[\left(3\chi+\bar\chi\right)
\left(\Qx\R^-_{20}+\Q\R^-_{00}/\sqrt2\right)
+\left(3\chi-\bar\chi\right)
\left(\bQx\bar\R^-_{20}+\bQ\bar\R^-_{00}/\sqrt2\right)\right]
,\nonumber\\
     \dot{\R}^-_{20}&=&
\frac{2}{m} \L^-_{20},
\nonumber\\
\dot {\P}^{-}_{20}&=&
-2\left[m\,\omega^2 + 2\kappa_0\Q\right] \L^-_{20} +4\sqrt2\kappa_0\Q\L^-_{00} 
- 4\alpha\kappa_0\bQ\left[\bar\L^-_{20}-\sqrt2 \bar\L^-_{00}\right]
\nonumber\\&&
+\frac{3I_2\hbar^2}{2A_1A_2}\chi\left[\Q\L^-_{00}/\sqrt2+\Qx\L^-_{20}+\bQ\bar\L^-_{00}/\sqrt2+\bQx\bar\L^-_{20}\right]
,\nonumber\\
     \dot{\L}^{-}_{00}&=&
\frac{1}{m} \P^-_{00}
- m\,\omega^2 \R^-_{00} + 2\sqrt2\kappa_0\left[\Q \R^-_{20} + \alpha\bQ \bar\R^-_{20}\right]\nonumber\\&&\qquad\quad
+\frac{\hbar^2}{12 A_1A_2}
\left[\left(\left(3\chi+\bar\chi\right)I_1-27\chi I_2/4\right)
\left(\left(2\Qx+\Q\right)\R^-_{00}+\sqrt2\Q\R^-_{20}\right)\right.
\nonumber\\&&\qquad\qquad\qquad\quad
\left.+\left(\left(3\chi-\bar\chi\right)I_1-27\chi I_2/4\right)
\left(\left(2\bQx+\bQ\right)\bar\R^-_{00}+\sqrt2\bQ\bar\R^-_{20}\right)\right]
,\nonumber\\
     \dot{\R}^-_{00}&=&
\frac{2}{m} \L^-_{00},
\nonumber\\
\dot{\P}^{-}_{00}&=&
-2m\,\omega^2 \L^-_{00} + 4\sqrt2\kappa_0\left[\Q \L^-_{20}+\alpha\bQ \bar\L^-_{20}\right]
\nonumber\\&&
+\frac{3I_2\hbar^2}{2A_1A_2}\chi\left[\Q\L^-_{20}/\sqrt2+\left(\Qx+\Q/2\right)\L^-_{00} + 
\bQ\bar\L^-_{20}/\sqrt2+\left(\bQx+\bQ/2\right)\bar\L^-_{00}\right],
\nonumber
\end{eqnarray}\begin{eqnarray}
 \dot{{\tilde \L}}_{20} &=& -\frac{\Delta}{2\hbar}  \L^-_{20}-\frac{\bar\Delta}{2\hbar} \bar \L^-_{20},
\nonumber\\
     \dot{\L}^{-}_{10}&=&\frac{4}{\hbar} \left[I_{rp}^{\kappa\Delta} {{\tilde \L}}_{10}
+\bar I_{rp}^{\kappa\Delta} {\bar{\tilde \L}}_{10}\right],
\nonumber\\
 \dot{{\tilde \L}}_{10} &=& -\frac{\Delta}{2\hbar}  \L^-_{10}-\frac{\bar\Delta}{2\hbar} \bar \L^-_{10}.
\end{eqnarray}

The following notations are used in (\ref{IV_mu=2})-(\ref{IS_mu=0}): 
\begin{eqnarray}
\label{Ai}
A_1=\sqrt2\, R_{20}^\eq-R_{00}^\eq=\frac{Q_{00}}{\sqrt3}\left(1+\frac{4}{3}\delta\right),\
A_2= R_{20}^\eq/\sqrt2+R_{00}^\eq=-\frac{Q_{00}}{\sqrt3}\left(1-\frac{2}{3}\delta\right),\qquad 
\end{eqnarray}
$Q_{00}=\frac{3}{5}AR_0^2/
[(1+\frac{4}{3}\delta)^{1/3}(1-\frac{2}{3}\delta)^{2/3}],$
\quad $\delta$ -- deformation parameter,\quad $Q_{20}=\frac{4}{3}\delta Q_{00},$\\
$\bar Q_{00}=Q_{00}^{\rm n}-Q_{00}^{\rm p},$ \,
$\bar Q_{20}=Q_{20}^{\rm n}-Q_{20}^{\rm p},$ \quad
$\bar \Delta=\Delta^{\rm n}-\Delta^{\rm p},$ \quad
$\Delta=\Delta^{\rm n}+\Delta^{\rm p},$ \,\,$R_0=r_0A^{1/3},$ \,
$r_0=1.2$~fm,
$$\displaystyle I_1=\frac{\pi}{4}\int\limits_{0}^{\infty}dr\, r^4\left(\frac{\partial n(r)}{\partial r}\right)^2,\quad
I_2=\frac{\pi}{4}\int\limits_{0}^{\infty}dr\, r^2 n(r)^2,\quad
K_0^{\tau}=\int d(\br,\bp) \kappa_0^{\tau}(\br,\bp),$$
$\bar K_0=K_0^{\rm n}-K_0^{\rm p},\quad
 K_0=K_0^{\rm n}+K_0^{\rm p},$\quad
$n(r)=n_0\left(1+{\rm e}^{\frac{r-R_0}{a}}\right)^{-1}$ -- nuclear density, 
$a=0.53$ fm;
$g=(g_s^{\rm n}+g_s^{\rm p})/4$.
The spin-orbit strength constant $\kappa_{\rm Nils}=0.0637$, $\eta=2\omega_0\kappa_{\rm Nils}$.

Anomalous density and semiclassical gap equation \cite{Ring}:
\begin{eqnarray}
&&\kappa(\br,\bp)=\frac{1}{2}
\frac{\Delta(\br,\bp)}{\sqrt{h^2(\br,\bp)+\Delta^2(\br,\bp)}},
\\
&&\Delta(\br,\bp)=-\frac{1}{2}\int\!\frac{d^3\!p'}{(2\pi\hbar)^3}
v(|\bp-\bp'|)
\frac{\Delta(\br,\bp')}{\sqrt{h^2(\br,\bp')+\Delta^2(\br,\bp')}},
\label{end}
\end{eqnarray}
where $v(|\bp-\bp'|)=\beta {\rm e}^{-\alpha |\bp-\bp'|^2}\!$
with $\beta=-|V_0|(r_p\sqrt{\pi})^3$ and $\alpha=r_p^2/4\hbar^2$.\\
$\di I^{\kappa\Delta}_{pp}(\br,p)=|V_0|
\frac{r_p^3}{\sqrt{\pi}\hbar^3}{\rm e}^{-\alpha p^2}
\int\!\kappa^r(\br,p')\left[\phi_0(x)
-4\alpha^2p'^4\phi_2(x)\right]
{\rm e}^{-\alpha p'^2}p'^2dp'$,\\
$\di I^{\kappa\Delta}_{rp}(\br,p)=|V_0|
\frac{r_p^3}{\sqrt{\pi}\hbar^3}{\rm e}^{-\alpha p^2}
\int\!\kappa^r(\br,p')[\phi_0(x)
-2\alpha p'^2\phi_1(x)]{\rm e}^{-\alpha p'^2}p'^2dp'$,\\
where
$$\phi_0(x)=\frac{1}{x}\sinh(x),\
\phi_1(x)=\frac{1}{x^2}\left[\cosh(x)-\frac{1}{x}\sinh(x)\right],$$
$$\phi_2(x)=\frac{1}{x^3}\left[\left(1
+\frac{3}{x^2}\right)\sinh(x)-\frac{3}{x}\cosh(x)\right]\ \mbox{\rm with}\ x=2\alpha pp'.$$
%
Parameters of pair correlations for WFM calculations: $V_0^{\rm p}=27$~MeV, $V_0^{\rm n}=23$~MeV, 
$r_p^{\rm p}=1.50$~fm, $r_p^{\rm n}=1.85$~fm for nuclei with $A=150-186$.

\section{Excitation probabilities }
\label{AppB}
   
Excitation
probabilities are calculated with the help of the theory of linear 
response of the system to a weak external field
\begin{equation}
\label{extf}
\hat O(t)=\hat O \,\e^{-i\Omega t}+\hat O^{\dagger}\,e^{i\Omega t}.
\end{equation}
A detailed explanation can be found in \cite{BaMo,BaSc}. 
We recall only the main points.
The matrix elements of the operator $\hat O$ obey the relationship \cite{Lane}
\begin{equation}
\label{matel}
|\langle\psi_a|\hat O|\psi_0\rangle|^2=
\hbar\lim_{\Omega\to\Omega_a}(\Omega-\Omega_a)
\overline{\langle\psi'|\hat O|\psi'\rangle\e^{-i\Omega t}},
\end{equation}
where $\psi_0$ and $\psi_a$ are the stationary wave functions of the
unperturbed ground and excited states; $\psi'$ is the wave function
of the perturbed ground state, $\Omega_a=(E_a-E_0)/\hbar$ are the
normal frequencies, the bar means averaging over a time interval much
larger than $1/\Omega$.

To calculate the magnetic transition probability, it is necessary
to excite the system by the following external field:
\begin{equation}
\label{Magnet}
\hat O_{\lambda\mu}=\mu_N\left(g_s^{\tau}\hat\bS/\hbar-ig_l^{\tau}\frac{2}{\lambda+1}[\br\times\nabla]\right)
\nabla(r^{\lambda}Y_{\lambda\mu}), \quad 
\mu_N=\frac{e\hbar}{2mc}.
\end{equation}
The free particle $g$-factors are given by
$g_l^{\rm p}=1,$ $g_s^{\rm p\, free}=5.5856$ for protons and $g_l^{\rm n}=0,$ $g_s^{\rm n\, free}=-3.8263$ for neutrons.
The spin quenching factor $q=0.7$ was applied in all our calculations: $g_s^\tau=q g_s^{\tau\rm\, free}$. 
The dipole operator \mbox{($\lambda=1,\ \mu=1$)} 
in cyclic coordinates looks like
\begin{equation}
\label{Magnet_1}
\hat O_{11}=
\sqrt{\frac{3}{4\pi}}\left[g_s^{\tau}\hat S_{1}/\hbar
-g_l^{\tau}\sqrt2\sum_{\nu,\sigma}
C_{1\nu,1\sigma}^{11}r_{\nu}\nabla_{\sigma}\right]\mu_N.
\end{equation}
 For the matrix element we have
\begin{eqnarray}
\label{psiO}
\langle\psi'|\hat O_{11}|\psi'\rangle &=&
\sqrt{\frac{3}{2\pi}}\left[-\frac{\hbar}{2}
(g_s^{\rm n}\F^{\rm n\downarrow\uparrow}
+g_s^{\rm p}\F^{\rm p\downarrow\uparrow})
-ig_l^{\rm p}\L_{11}^{\rm p+}\right]\frac{\mu_N}{\hbar}
\nonumber\\
&=&\sqrt{\frac{3}{8\pi}}\left[
\frac{1}{2}(g_s^{\rm p}-g_s^{\rm n})\bar{\F}^{\downarrow\uparrow}
+\frac{i}{\hbar}g_l^{\rm p} \bar{\L}_{11}^{+}
+\frac{i}{\hbar}[g_s^{\rm n}+g_s^{\rm p}-g_l^{\rm p}]\L_{11}^{+}
\right]\mu_N.\qquad
\end{eqnarray}
 Deriving (\ref{psiO}) we have used the relation $2i\L^+_{11}=-\hbar \F^{\d}$,
which follows from the angular momentum conservation~(\ref{J1}).
One has to add the external field (\ref{Magnet_1}) to the Hamiltonian (\ref{Ham}). 
Due to the external field some dynamical equations of 
(\ref{iv}) become inhomogeneous:
\begin{eqnarray}
     \dot {\bar\R}^{+}_{21}
&=&\ldots\; +i\frac{3}{\sqrt{\pi}}\frac{\mu_N}{2\hbar} g^{\rm p}_l R^{{\rm p}+}_{20}(\eq)\,\e^{i\Omega t},
\nonumber\\
     \dot {\bar\L}^{-}_{11}
&=&\ldots\; + i\sqrt{\frac{3}{\pi}}\frac{\mu_N}{2\hbar} g^{\rm p}_l L^{{\rm p}-}_{10}(\eq)\,\e^{i\Omega t},
\nonumber\\
     \dot {\bar\L}^{\d}_{10}
&=&\ldots\; 
+i\sqrt{\frac{3}{2\pi}}\frac{\mu_N}{2\hbar}\left[ g^{\rm n}_s L^{{\rm n}-}_{10}(\eq)-g^{\rm p}_s L^{{\rm p}-}_{10}(\eq)\right]  \e^{i\Omega t}.
\end{eqnarray}
For the isoscalar set of equations (\ref{is}), respectively, we obtain:
\begin{eqnarray}
     \dot {\R}^{+}_{21}
&=&\ldots\; - i\frac{3}{\sqrt{\pi}}\frac{\mu_N}{2\hbar} g^{\rm p}_l R^{{\rm p}+}_{20}(\eq)\,\e^{i\Omega t},
\nonumber\\
     \dot {\L}^{-}_{11}
&=&\ldots\; - i\sqrt{\frac{3}{\pi}}\frac{\mu_N}{2\hbar} g^{\rm p}_l L^{{\rm p}-}_{10}(\eq)\,\e^{i\Omega t},
\nonumber\\
     \dot {\L}^{\d}_{10}
&=&\ldots\; 
+i\sqrt{\frac{3}{2\pi}}\frac{\mu_N}{2\hbar}\left[ g^{\rm n}_s L^{{\rm n}-}_{10}(\eq)+g^{\rm p}_s L^{{\rm p}-}_{10}(\eq)\right] \e^{i\Omega t}.
\end{eqnarray}
Solving the inhomogeneous set of equations 
one can find the required 
in (\ref{psiO}) 
values of
$\L_{11}^{+}$ , $\bar\L_{11}^{+}$ and $\bar\F^{\d}$ and  calculate 
$B(M1)$ factors for all excitations as it is explained in \cite{BaSc,BaMo}.


To calculate the electric transition probability, it is 
necessary to excite the system by the external field operator
\begin{equation}
\label{O2mu}
\hat O_{2\mu}=er^2Y_{2\mu}=\beta \{r\otimes r\}_{2\mu},
\end{equation}
where $\beta=e\sqrt\frac{15}{8\pi}$.
 The matrix element is given by
\begin{equation}
\label{psiO2}
\langle\psi'|\hat O_{2\mu}|\psi'\rangle= \beta\,\R_{2\mu}^{\rm p+}
=\frac{\beta}{2} (\R_{2\mu}^+-\bar \R_{2\mu}^+).
\end{equation}
 For $\mu=1$ the external field makes inhomogeneous the following equations from (\ref{iv}), (\ref{is}):
\begin{eqnarray}
     \dot{\bar\L}^{+}_{21}&=&\ldots\; 
+\frac{2}{3}\beta\left(Q_{00}^{\rm p}+\frac{Q_{20}^{\rm p}}{4}\right)
\e^{i\Omega t},
\nonumber\\
     \dot{\bar\L}^{+}_{11}&=&\ldots\; 
-\frac{\beta}{2}Q_{20}^{\rm p}\e^{i\Omega t},
\nonumber\\
     \dot{\bar\P}^{-}_{21}&=&\ldots\; 
+\sqrt{2}\beta L^{{\rm p}-}_{10}(\eq)\e^{i\Omega t},
\end{eqnarray}
\begin{eqnarray}
     \dot \L^{+}_{21}&=&\ldots\; 
-\frac{2}{3}\beta\left(Q_{00}^{\rm p}+\frac{Q_{20}^{\rm p}}{4}\right)
\e^{i\Omega t},
\nonumber\\
     \dot \L^{+}_{11}&=&\ldots\; 
+\frac{\beta}{2}Q_{20}^{\rm p}\e^{i\Omega t},
\nonumber\\
     \dot \P^{-}_{21}&=&\ldots\; 
-\sqrt{2}\beta L^{{\rm p}-}_{10}(\eq)\e^{i\Omega t}.
\end{eqnarray}

\section{Angular momentum conservation}\label{AppC}

It was shown~\cite{BaMo} that the average value of the $\mu=1$ component of the orbital moment $\langle\hat l_1\rangle=-i\sqrt2 L^+_{11}(t)$. 
The  corresponding average value of the spin operator is $\langle\hat s_1\rangle=-\frac{\hbar}{\sqrt2} F^\d(t)$. 
It is also easy to verify that 
$\langle\hat l_{-1}\rangle=-i\sqrt2 L^+_{1-1}(t)$ and $\langle\hat l_0\rangle=-i\sqrt2 L^+_{10}(t)$. 
For spin operator we obtain: $\langle\hat s_{-1}\rangle=\frac{\hbar}{\sqrt2} F^\u(t)$ and $\langle\hat s_0\rangle=-\frac{\hbar}{2} F^-(t)$.
The total angular momentum $\langle\hat{\bf J}\rangle=\langle\hat {\bf l}\rangle +\langle\hat {\bf s}\rangle$ can be written in terms of dynamical variables:
\begin{eqnarray}
 &&\langle\hat J_{-1}\rangle=-i\sqrt2L^+_{1-1}+\frac{\hbar}{\sqrt2} F^\u,
 \\
 &&\langle\hat J_0\rangle=-i\sqrt2L^+_{10}+\frac{\hbar}{2} F^-,
 \\
 &&\langle\hat J_1\rangle=-i\sqrt2L^+_{11}-\frac{\hbar}{\sqrt2} F^\d. \label{J1}
 \end{eqnarray}
It is easy to see that such combinations of the respective isoscalar dynamical equations from Appendix~\ref{AppA} are equal to zero, 
that means that the total angular momentum is conserved.

\end{widetext}

\end{document}